\documentclass[aps,twocolumn,superscriptaddress]{revtex4-1}
\usepackage{inputenc}
\usepackage{amssymb}
\usepackage{amsmath}
\usepackage{amsfonts}
\usepackage{latexsym}
\usepackage{graphicx}        
\usepackage{graphics}        

\def\Cas{\mathrm{Cas}}
\def\PFA{\mathrm{PFA}}
\def\P{\mathrm{P}}    
\def\D{\mathrm{D}}    
\def\B{\mathrm{B}}    
\def\T{\mathrm{T}}    
\def\dd{\mathrm{d}}   
\def\imag{{\imath}}   
\def\max{\mathrm{max}}
\def\min{\mathrm{min}}
\def\iin{\mathrm{in}}
\def\out{\mathrm{out}}
\def\cav{\mathrm{cav}}
\def\L{\mathrm{L}}
\def\R{\mathrm{R}}
\def\C{\mathrm{C}}
\def\patch{\mathrm{patch}}
\def\rms{\mathrm{rms}}
\def\bk{\mathbf{k}}
\def\br{\mathbf{r}}
\def\Re{\mathop{\rm Re}}
\def\Im{\mathop{\rm Im}}
\def\Tr{\mathop{\rm Tr}}

\def\calH{\mathcal{H}}
\def\calF{\mathcal{F}}
\def\calL{\mathcal{L}}

\def\calO{\mathcal{O}}
\def\calD{\mathcal{D}}
\def\calK{\mathcal{K}}
\def\calR{\mathcal{R}}
\def\ovn{\overline{n}}

\def\ovE{\overline{E}}

\begin{document}

\title{Casimir forces}

\author{S. Reynaud}
\affiliation{Laboratoire Kastler Brossel, CNRS, ENS-PSL Research
University, Coll\`ege de France, UPMC-Sorbonne Universit\'es,
Campus Jussieu, F-75252 Paris, France.}
\author{A. Lambrecht}
\affiliation{Laboratoire Kastler Brossel, CNRS, ENS-PSL Research
University, Coll\`ege de France, UPMC-Sorbonne Universit\'es,
Campus Jussieu, F-75252 Paris, France.}
\date{\today, to appear in \emph{Quantum Optics and Nanophotonics},
Oxford University Press}

\begin{abstract}
The present notes are organized as the lectures given at the Les
Houches Summer School ``Quantum Optics and Nanophotonics'' in August
2013. The first section contains an introduction and a description
of the current state-of-the-art for Casimir force measurements and
their comparison with theory. The second and third sections are a
pedagogical presentation of the main features of the theory of
Casimir forces for 1-dimensional model systems and for mirrors in
3-dimensional space.
\end{abstract}

\maketitle


\subsection*{Introduction}

The emergence of quantum theory has profoundly altered our
conception of space by forcing us to consider it as permanently
filled by vacuum field fluctuations~\cite{Milonni1994,Milton2001}.
These vacuum fluctuations are electromagnetic fields propagating
with the speed of light, as any free field, and corresponding to an
energy of half a photon per mode. They have a number of observable
consequences in microscopic physics, for example the radiative
corrections in subatomic physics~\cite{Itzykson1985}, the
spontaneous emission processes, the Casimir-Polder interaction and
Lamb shift in atomic physics~\cite{Cohen-Tannoudji1992}.

Vacuum fluctuations also have observable mechanical effects in
macroscopic physics and the archetype of these effects is the
Casimir force between two motionless mirrors. This force was
predicted in 1948 by H.~B.~G. Casimir~\cite{Casimir1948} and soon
observed in experiments~\cite{Lamoreaux1999}. Experiments have been
improved after years of development and they now reach a good level of
precision~\cite{Decca2003,Decca2005ap,Decca2007prd,Decca2007epj}
(more references below). However, the comparison with theoretical
predictions has raised difficulties which have been discussed in a
large number of papers (references in~\cite{Klimchitskaya2009,Dalvit2011}).

This comparison is particularly interesting because of its
fascinating interfaces with open questions in fundamental physics.
The Casimir effect is connected with the puzzles of gravitational
physics through the problem of vacuum energy as well as with the
principle of relativity of motion~\cite{Jaekel1997}. Effects beyond
the Proximity Force Approximation also make apparent the rich
interplay of vacuum and geometry~\cite{Balian2003}. Casimir physics
also plays an important role in the tests of gravity at
sub-millimeter
ranges~\cite{Fischbach1998,Adelberger2003,Adelberger2009,Antoniadis2011}.
For scales of the order of the micrometer, such gravity tests are
performed by comparing Casimir force measurements with theory and
the comparison has to take into account the many differences between
real experiments and the idealized case considered initially by
Casimir~\cite{Lambrecht2003}.

At the end of this short general introduction, it has also to be
stressed that the Casimir and closely related Casimir-Polder
forces~\cite{Casimir1948pr,Feinberg1970,Power1993,Milton2011ajp,Intravaia2011}
have strong connections
with various active domains and interfaces of physics, such as
atomic and molecular physics, condensed matter and surface physics,
chemical and biological physics, micro- and
nano-technology~\cite{Parsegian2006,French2010}.

\subsection*{Outline}

The present paper is organized as the lectures given at the Les
Houches Summer School in August 2013~: the first section contains
an introduction and a description of the current state-of-the-art
for experiments and their comparison with theory while the second
and third sections provide a pedagogical presentation of the main
features of the theory of Casimir forces.

Section \ref{Lecture1} begins with a short history of quantum field
fluctuations in vacuum. We then review various arguments involved in
the comparison with theory of the experiments devoted to the
measurement of the Casimir force. As experiments are performed with
gold-covered plates, the force depends on non universal properties
of the real plates used in the experiments. As they are performed at
room temperature, the effect of thermal field fluctuations has to be
added to that of vacuum fluctuations. The most precise experiments
are performed in the plane-sphere geometry and not in the geometry
of two parallel planes whereas the latter is theoretically easier
to handle. Finally, surfaces are non ideal, and effects such as
roughness, electrostatic patches and contamination
affect the comparison between theory and experiment.

Section \ref{Lecture2} contains a simple derivation of the Casimir
effect in a model of scalar fields on a 1-dimensional line. This
model allows one to introduce the \emph{Quantum Optics} approach to
the Casimir effect. This approach is based on the existence of
field fluctuations which pervade empty space and exert
\emph{radiation pressure} on mirrors at rest in vacuum. The force is thus
calculated as the result of different pressures acting on inner and
outer sides of the two mirrors which form a cavity. This approach is
often also called the \emph{Scattering Formalism} because all
properties of the Casimir force are determined by the reflection
amplitudes of the fields on the two mirrors~\cite{Jaekel1991}.

Section \ref{Lecture3} then treats the case of two plane and
parallel mirrors at rest in electromagnetic vacuum in 3-dimensional
space. It describes the models generally used for the metallic
mirrors used in the experiments~\cite{Lambrecht2000,Genet2003} and
discusses the results obtained in this manner. For mirrors
characterized by Fresnel reflection amplitudes deduced from a linear
and local dielectric function, the Scattering Formalism leads to the
same results as Lifshitz's
method~\cite{Lifshitz1956,Dzyaloshinskii1961,Lifshitz1980}. The section ends up
with a presentation of the general scattering formalism which allows
one to deal with non specular reflection and arbitrary
geometries~\cite{Lambrecht2006,Emig2007,Milton2008,Lambrecht2011,Rahi2011,Rodriguez2011}

\section{Vacuum fluctuations and Casimir forces}
\label{Lecture1}

\subsection*{The birth of quantum vacuum}

The classical idealization of absolute empty space was affected by
the discovery of black body radiation which is present everywhere at
non zero temperature and exerts a pressure onto the boundaries of
any cavity. It is precisely for explaining the properties of black
body radiation that Planck introduced the first quantum law in
1900~\cite{Planck1900} (discussions in~\cite{Darrigol2000}).

In modern terms, the first Planck law gives the energy per
electromagnetic mode characterized by its frequency $\omega = 2 \pi
\nu$ as the product of the energy of a single photon $\hbar \omega
\equiv h \nu $ by a mean number of photons $\ovn$ per mode
\begin{equation}
\ovE_{1900} = \ovn \hbar \omega  \quad, \qquad \ovn \equiv \frac
{1}{e^{ \hbar \omega/k_\B  T } - 1}~. \label{Planck1900}
\end{equation}
Unsatisfied with his first derivation, Planck~\cite{Planck1911}
wrote in 1911 a new expression for the mean energy per mode
\begin{equation}
\ovE_{1911} = \left( \frac 12 + \ovn \right) \hbar \omega~.
\label{Planck1911}
\end{equation}
The difference between the two Planck laws corresponds to the
zero-point energy. Whereas the first law describes a cavity entirely
emptied out of radiation at the limit of zero temperature
($\ovn\to0$ when $T\to0$), the zero-point energy added in the second
law persists at zero temperature.

The story of the two Planck laws and of the discussions they raised
is related for example in~\cite{Milonni1991}. The arguments used by
Planck in 1911 are no longer considered as satisfactory nowadays but
an argument which is still valid was proposed by Einstein and Stern
in 1913~\cite{Einstein1913}. In order to state this argument, let us
consider the limit for $\ovE$ at high temperatures ($k_\B T\gg\hbar
\omega$)
\begin{eqnarray}
\ovE_{1900} &=& \frac {\hbar \omega}{e^{ \hbar \omega/k_\B  T } - 1}
= k_\B T  - \frac{\hbar \omega}2 + \frac{\left(\hbar \omega
\right)^2}{4k_\B T} + \ldots \nonumber\\
\ovE_{1911} &=& \frac{\hbar \omega}{2} + \ovE_{1900} = k_\B T +
\calO \left( \frac{\left(\hbar \omega \right)^2}{k_\B T} \right) ~.
\end{eqnarray}
In contrast to the first Planck's law \eqref{Planck1900} which falls
off the correct classical limit by a constant offset $\frac 12 \hbar
\omega$, the second Planck's law \eqref{Planck1911} matches the
correct classical limit at high temperatures. This feature is
emphasized by the modern writing of this law which has clearly no
term linear in $\hbar\omega$ because the right-hand
side is an even function of $\omega$ (note that
$\coth(x) \equiv 1/\tanh (x)$)
\begin{equation}
\ovE_{1911} = \left( \frac 12 + \ovn \right) \hbar \omega = \frac
{\hbar \omega}{2} \coth \frac{\hbar\omega}{2k_\B T} ~. \label{Planckmodern}
\end{equation}

Debye was the first to insist on observable consequences of
zero-point fluctuations in atomic motion, by discussing their effect
on the intensities of diffraction peaks~\cite{Debye1914} whereas
Mulliken gave the first experimental proof of these consequences by
studying vibrational spectra of molecules~\cite{Mulliken1924}. At
this point, we may emphasize that these discussions took place
before the existence of these fluctuations was confirmed by fully
consistent quantum theoretical calculations. Nowadays, vacuum
fluctuations are just an immediate consequence of Heisenberg
inequalities (see for example references
in~\cite{Reynaud1990,Reynaud1992,Reynaud1997}).

\subsection*{The puzzle of vacuum energy}

Nernst is credited for having been the first to emphasize in 1916
that zero-point fluctuations should also exist for free
electromagnetic fields~\cite{Nernst1916} (discussions
in~\cite{Browne1995}), thus discovering what physicists now call
\emph{quantum vacuum}. He also noticed that the associated energy
constituted a challenge for gravitation theory. When summing up the
zero-point energies over all field modes, a finite energy density is
obtained for the first Planck law - this is the solution of the
\emph{ultraviolet catastrophe} - but an infinite value is produced
from the second law. When introducing a high frequency cutoff, the
calculated energy density remains finite but it is still much larger
than the mean energy observed in the world around us through
gravitational phenomena~\cite{Weinberg1989}. Pedagogical derivations
and numbers illustrating this major problem are given
in~\cite{Adler1995}.

This puzzle has led famous physicists to deny the reality of vacuum
fluctuations. In particular, Pauli stated in his textbook on Wave
Mechanics~\cite{Pauli1933} (translation from~\cite{Enz1974}):
\emph{At this point it should be noted that it is more consistent
here, in contrast to the material oscillator, not to introduce a
zero-point energy of $\frac12\hbar\omega$ per degree of freedom.
For, on the one hand, the latter would give rise to an infinitely
large energy per unit volume due to the infinite number of degrees
of freedom, on the other hand, it would be principally unobservable
since nor can it be emitted, absorbed or scattered and hence, cannot
be contained within walls and, as is evident from experience,
neither does it produce any gravitational field.}

A part of these statements is simply unescapable: it is just a
matter of evidence that the mean value of vacuum energy as predicted
by quantum theory does not contribute to gravitation as an ordinary
energy. However, we know nowadays that vacuum fluctuations are
\emph{scattered} by matter, as shown by the numerous effects of the
associated scattering in subatomic~\cite{Itzykson1985} and
atomic~\cite{Cohen-Tannoudji1992} physics. And the Casimir effect
discussed in the sequel of this paper may be seen as the
manifestation of vacuum fluctuations when being \emph{contained
within walls}. Let us note at this point that different points of
view coexist about the significance of vacuum
fluctuations~\cite{Enz1974,Schwinger1975,Aitchinson1985,Saunders1991,%
Sciama1991,Rugh2002,Jaffe2005,Kragh2012ahes,Kragh2012aag}.

The puzzle of vacuum energy has been discovered nearly one century
ago and it is not yet solved. It has led and still leads to many
ideas: for example, a \emph{dark energy length scale}
$\lambda$=85$\mu$m can be defined by setting the cutoff used to
calculate the vacuum energy so that it fits the now measured cosmic
vacuum energy density~\cite{Riess1998,Perlmutter1999,Bennett2003}.
It is thus natural to check if gravity could be affected below this
dark energy length scale~\cite{Beane1997,Dvali2001,Sundrum2004}. It
has been shown in torsion-balance experiments~\cite{Kapner2007} that
Yukawa modifications of the gravitational inverse-square law can
have a magnitude equal to that of gravity only if their range has a
value smaller than 56$\mu$m, which discards this kind of ideas under
their simplest forms. More possibilities, corresponding for example
to power-law modifications associated with compact
extra-dimensions~\cite{Arkani-Hamed1998}, are discussed
in~\cite{Adelberger2003,Adelberger2009,Antoniadis2011}.

The search for scale-dependent modifications of the gravity force
law are currently pushed down to even smaller ranges and they
approach the micrometer distance range where Casimir forces are
predominant. For these \emph{Casimir tests} of the gravity law to
make sense, the accuracy and reliability of theoretical and
experimental values have to be assessed cautiously and independently
of each other. In particular, systematic effects have to be
identified and eliminated, whenever this is possible. This implies in
particular to deal carefully with the many differences between the
idealized situation studied by Casimir and the configuration of real
experiments~\cite{Lambrecht2012}.

\subsection*{The Casimir force between ideal and real mirrors}

\begin{figure}[tbh]
\includegraphics[width=0.25\textwidth]{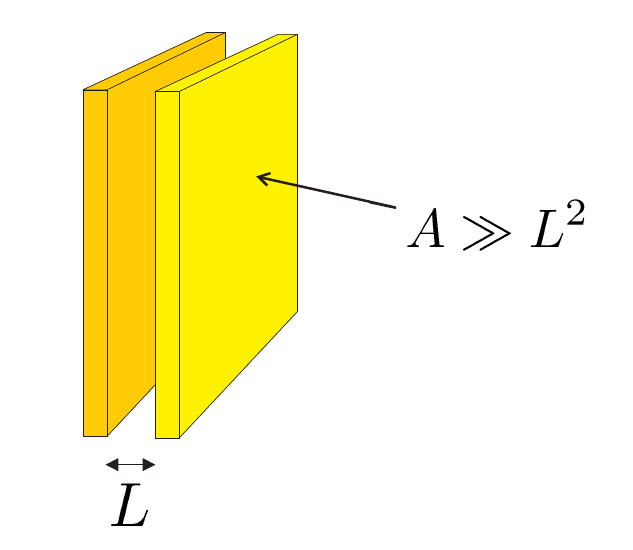}
\caption{Configuration considered by Casimir~:
two perfectly parallel planes placed in vacuum
experience an attractive pressure given
by~\eqref{CasimirIdealP} under the assumptions
of perfect reflection and zero temperature.}
\label{FigPlanePlane}
\end{figure}

In his initial calculation~\cite{Casimir1948}, Casimir considered an
idealized configuration, with perfectly smooth, flat and parallel
plates (see Figure \ref{FigPlanePlane}) in the limit of zero
temperature and perfect reflection. $L$ denotes the distance between
the two plates and $A$ their area, supposed to be large enough
($A\gg L^2$).

He thus obtained expressions for the energy $E_\Cas$ and force
$F_\Cas \equiv - \dd E_\Cas/\dd L$ which exhibit a universal
behavior associated with the confinement of vacuum energy
\begin{eqnarray}
E_\Cas &=& - \frac{\hbar c \pi^2 A}{720 L^3} \quad,\quad F_\Cas  = -
\frac{\hbar c \pi^2 A}{240 L^4}~, \label{CasimirIdeal}
\end{eqnarray}
where $c$ is the speed of light and
$\hbar=h/2\pi$ the reduced Planck constant. The signs have been chosen
according to the standard thermodynamical conventions (the relation
with thermodynamics of the Casimir effect will be discussed later on).
The negative energy
corresponds to a binding energy and the negative force to an
attractive force, that is also a negative pressure
\begin{eqnarray}
P_\Cas  &\equiv& \frac{F_\Cas}A = - \frac{\hbar c \pi^2} {240 L^4}~.
\label{CasimirIdealP}
\end{eqnarray}

The order of magnitude of the pressure is
$\left|P_\Cas\right|\sim$1mPa for two mirrors at the distance
$L=1\mu$m typical for Casimir force measurements (see below). The
formula \eqref{CasimirIdealP} describes an extremely rapid increase
of the pressure when the distance is decreased, and it would lead
to a value $\sim$1TPa typical of strong molecular cohesion
when it is extrapolated down to atomic distances $\sim$0.1nm. This
means that the Casimir force is a quantum force like molecular
cohesion forces, which has a weaker magnitude only because it is
measured at distances much larger than typical atomic distances.
Note however that formula \eqref{CasimirIdealP} cannot be used
at atomic distances where the ideal assumptions used to derive it
are no longer valid, as we explain now.

\subsection*{The effect of imperfect reflection}

Indeed, perfectly reflecting mirrors do not exist, except as
idealizations giving fair descriptions of reality in limiting
cases only. The mirrors used in
the experiments are made of metal and they have a good reflection
only at frequencies below the plasma frequency. Accounting for their
imperfect reflection and its frequency dependence is thus essential
for obtaining a reliable theoretical expectation of the Casimir
pressure~\cite{Lambrecht2003}. In other words,
the real Casimir pressure depends on the non universal optical
properties of the material plates used in the experiments. It can be
written as the product of the ideal result \eqref{CasimirIdealP} by
a dimensionless factor which accounts for these optical properties
\begin{eqnarray}
P = P_\Cas \eta_\P~. \label{CasimirRealP}
\end{eqnarray}
The expression of $\eta_\P$ in terms of the optical properties
of the mirrors will be given in section~\ref{Lecture3}.

Most descriptions of the metallic mirrors used in the experiments
are based on Fresnel reflection laws at the two interfaces between
vacuum and metallic bulks with optical properties
described by a linear and local dielectric response function.
This dielectric function
$\varepsilon$ is a sum of contributions corresponding respectively
to bound ($\bar{\varepsilon}$) and conduction electrons, the latter
being directly related to the conductivity ($\sigma$)
\begin{eqnarray}
\varepsilon[\omega] = \bar{\varepsilon}[\omega] +
\frac{\sigma[\omega]}{-\imag \omega} ~.\label{dielectricfunction}
\end{eqnarray}
Note that functions $\varepsilon$, $\bar{\varepsilon}$ and $\sigma$
are all defined as reduced quantities, with their SI counterparts
being $\varepsilon_0\varepsilon$, $\varepsilon_0\bar{\varepsilon}$
and $\varepsilon_0\sigma$ ($\varepsilon_0=1/\mu_0 c^2$ is the vacuum
permittivity). With these conventions, $\varepsilon$ and
$\bar{\varepsilon}$ are dimensionless while $\sigma$ has the
dimension of a frequency.

The dielectric function \eqref{dielectricfunction} has to be
obtained from optical data~\cite{Lambrecht2000,Svetovoy2008prb} as
they are tabulated for example for gold in~\cite{Palik1995}. At low
frequencies, $\bar{\varepsilon}$ tends to a constant while the
contribution of conduction electrons diverges while $\sigma$ tends to a
constant $\sigma_0$. Optical data have then to be extrapolated at
low frequencies by using the dissipative Drude model for the
conductivity of the metal~\cite{Ashcroft1976}
\begin{eqnarray}
\sigma_\D [\omega] = \frac{\omega_\P^2}{\gamma-\imag \omega}~.
\label{Drudemodel}
\end{eqnarray}
Here $\omega_\P$ is the plasma frequency and $\gamma$ the relaxation
parameter for conduction electrons. This model meets the well-known
fact that gold has a finite static conductivity
\begin{eqnarray}
\sigma_0 = \frac{\omega_\P^2}\gamma~. \label{finiteconductivity}
\end{eqnarray}

For reasons which will become clear in the following, the limiting case
of a lossless plasma of conduction electrons ($\gamma = 0$ in
\eqref{Drudemodel}) is often considered
\begin{eqnarray}
\sigma_\P [\omega] = \frac{\omega_\P^2}{-\imag \omega}~.
\label{plasmamodel}
\end{eqnarray}
This so-called \emph{plasma model} cannot be an accurate description
of metallic mirrors. As a matter of fact, it contradicts the fact
that gold has a finite static conductivity
\eqref{finiteconductivity} while also leading to a poorer fit of
tabulated optical data than the more general Drude
model~\cite{Lambrecht2000}. However, $\gamma$ is much smaller than
$\omega_\P$ for a good metal (for example
$\gamma\simeq0.004\omega_\P$ for gold). As the difference between
\eqref{Drudemodel} and \eqref{plasmamodel} is appreciable only at
low frequencies $\omega\lesssim\gamma$ where $\varepsilon$ is very
large for both models, one might expect that it does not affect too
much the value of the Casimir force. This naive expectation is met
at small distances or low temperatures but not in the general case
of arbitrary distances and temperatures, as explained in the
following.

\subsection*{The effect of temperature}

Most experiments are performed at room temperature, so that the
effect of thermal fluctuations has to be added to that of vacuum
fields~\cite{Mehra1967,Brown1969,Schwinger1978,Genet2000}. This
important point will be discussed in a detailed manner in
sections~\ref{Lecture2} \& \ref{Lecture3}. At this point, we focus
on the strong correlation effect obtained between the effects of
temperature and dissipation, which has given rise to a large number
of contradictory papers (see for example the references
in~\cite{Milton2005,Klimchitskaya2006,Brevik2006,Ingold2009,Brevik2014}).

\begin{figure}[tbh]
\includegraphics[width=0.5\textwidth]{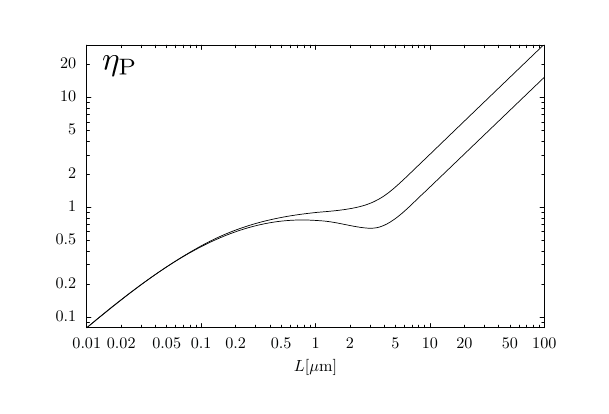}
\caption{Variation with distance $L$ of the Casimir pressure shown
as the ratio $\eta_\P$ of the real ($P$) to the ideal ($P_\Cas$)
Casimir pressure (see eq.\eqref{CasimirRealP}). $P$ is smaller than
$P_\Cas$ ($\eta_\P<1$) at small distances ($L\lesssim$1$\mu$m), due
to imperfect reflection, whereas it is larger ($\eta_\P>1$) at large
distances (1$\mu$m$\lesssim L$), due to the contribution of thermal
photons. In the latter large-distance domain, there is a significant
difference between the Drude (lower curve) and plasma (upper curve)
models, both drawn here for room temperature.} \label{FigThermal}
\end{figure}

Bostr\"{o}m and Sernelius~\cite{Bostrom2000} were the first to
remark that, in spite of the naive expectation described in the end
of the preceding subsection, dissipation has a large effect on the
value of the Casimir force at distances accessible in experiments
and at room temperature. Their result is illustrated on Figure
\ref{FigThermal} where the ratio $\eta_\P$ giving the real Casimir
pressure \eqref{CasimirRealP} is drawn for the Drude and plasma
models at room temperature ($T$=300K), using the formulas
in~\cite{Ingold2009}. These two models correspond to the dielectric
function \eqref{dielectricfunction} with $\bar{\varepsilon}$=1 and
$\sigma$ substituted by \eqref{Drudemodel} and \eqref{plasmamodel}
respectively. The simplification $\bar{\varepsilon}$=1 does not
change the difference between the predictions of the two models and
it can also be dropped, by having $\bar{\varepsilon}$ deduced from
the optical data. The parameters of the optical models are chosen to
match values typical for thick layers of gold ($\omega_\P=2\pi
c/\lambda_\P$ with $\lambda_\P$=136nm and $\gamma/\omega_\P=0.004$).

A striking difference appears between the predictions of the two
models~\cite{Bostrom2000}. These predictions, which are close to
each other at short distances, exhibit an increasing difference for
distances of the order or larger than 1$\mu$m. In particular, the
ratio of the plasma to Drude prediction for the Casimir pressure
goes to a factor 2 at the limit of large distances.
In fact, the result of the plasma
model coincides at this limit with that obtained for perfect mirrors
whereas the result of the Drude model reaches only half that value.
It is worth recalling here that this last result is reproduced by
the derivations of Casimir pressures from microscopic models of the
metallic mirrors~\cite{Jancovici2005,Buenzli2005,Bimonte2009}.

As a matter of principle, there should be no doubt that the Drude
model is a better representation of the optical properties of real
plates at low frequencies than the plasma model. At this point
however, we have to face discrepancies in the comparison between
experimental results and theoretical predictions and, unexpectedly,
some experimental results appear to lie closer to the predictions of
the plasma model than to that of the Drude
model~\cite{Klimchitskaya2009}. Before coming to this point, we have
still to discuss another important feature of the recent precise
experiments which are performed in the plane-sphere geometry and not
in the plane-plane geometry in which most calculations are done.

\subsection*{The effect of geometry}

\begin{figure}[tbh]
\includegraphics[width=0.25\textwidth]{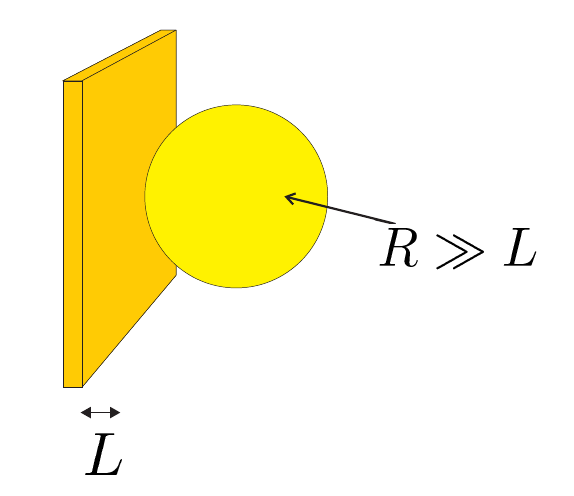}
\caption{Configuration for most Casimir experiments~:
a plane and a spherical mirror placed in vacuum
experience an attractive force.}
\label{FigPlaneSphere}
\end{figure}

The configuration used for most Casimir experiments corresponds to a
plane and a sphere, with $L$ the distance of closest approach and
$R$ the radius of the sphere, supposed to be large (see Figure
\ref{FigPlaneSphere}). The force in this plane-sphere geometry is
usually calculated by using the so-called \emph{Proximity Force
Approximation} (PFA)~\cite{Derjaguin1956,Decca2009}. Let us
emphasize here that this approximation is valid only at the limit of
a very large radius ($R\gg L$) and that the question of its accuracy
for a finite value of $R/L$ remains an open question (more
discussions on this topic later on).

We assume here provisionally that the PFA is precise enough for the
purpose of theory-experiment comparison. It follows that the force
between the plane and the sphere can be obtained by integrating over
the distribution of local inter-plate distances the pressure $P$
calculated in the geometry with two parallel planes. In the
plane-sphere geometry, this gives
\begin{eqnarray}
F_\PFA\left(L\right) = \int_L^\infty \dd A ~ P\left(\calL\right)
\quad,\quad \dd A = 2\pi R ~ \dd \calL~, \label{FPFA}
\end{eqnarray}
where $P\left(\calL\right)$ is the pressure evaluated between two
planes at at a distance $\calL$ from each other; $\calL$ runs
over distances larger than the distance of closest approach $L$
and $\dd A$ is the corresponding element of surface.

In the experiment, the gradient $G$ of the Casimir force in the
plane-sphere geometry is measured (see the next subsection). This
quantity is deduced from \eqref{FPFA} as
\begin{eqnarray}
G_\PFA\left(L\right) \equiv \frac{\partial F_\PFA}{\partial L}= -
2\pi R P\left(L\right)~. \label{GPFA}
\end{eqnarray}
Within PFA, this measurement gives the Casimir pressure evaluated at
distance $L$ between two planes, which can then be substituted by
the expression \eqref{CasimirRealP} (with $\eta_\P$ to be given in
section~\ref{Lecture3}).

\subsection*{Casimir experiments}

We now present briefly the experimental methods and results. To this
aim, we focus our attention on a few experiments~: the experiment at
IUPUI~\cite{Decca2005ap,Decca2007prd,Decca2007epj} which has been
run for ten years, with results pointing to an unexpected conclusion
later confirmed at UCR~\cite{Chang2012}, and the experiment in
Yale~\cite{Sushkov2011} which points to a different conclusion. We
also give here a list of other Casimir measurements
which have produced information of interest on the topics
discussed in this paper~\cite{Derjaguin1956,Sparnaay1958,%
Sabisky1973,Lamoreaux1997,Mohideen1998,Harris2000,Ederth2000,%
Chan2001,Bressi2002,Lisanti2005,Decca2005prl,Kim2008,%
Jourdan2009,Man2009,Masuda2009,Munday2009,Torricelli2010,%
Banishev2012,Banishev2013,Castillo-Garza2013}.

The experiment at IUPUI is described in the
papers~\cite{Decca2005ap,Decca2007prd,Decca2007epj}. A summary and
update can be found in the slides associated with a talk given
recently by R.S. Decca at a Pan-American Advanced Study Institute
school~\cite{Decca2012pasi}. The experiment uses dynamic
measurements of the resonance frequency of a torsion
micro-oscillator. For the free micro-oscillator, that is in the
absence of the Casimir force, the resonance frequency is determined
by the stiffness coefficient $K_0$ and the moment of inertia $I$
\begin{eqnarray}
\omega_0^2 = \frac{K_0}I~.
\end{eqnarray}
When a gold-covered sphere is approached from the gold-covered plane
of the micro-oscillator plate, the effective stiffness is modified
as the gradient of the Casimir force $G$. The resonance frequency is
thus shifted to a new value
\begin{eqnarray}
\omega^2 = \frac{K}I\quad,\quad K=K_0 - b^2 G~,
\end{eqnarray}
where $b$ is the lever arm.
As the radius of the sphere is $R\simeq150\mu$m and the range of
distances $0.16\mu$m$<L<0.75\mu$m, the condition $R\gg L$ is met.
Using the expression \eqref{GPFA} which gives the gradient $G$
within the PFA, and measuring accurately $b$, $I$ and $R$, the shift
of the squared frequency is then transformed into a reading of the
Casimir pressure $P\left(L\right)$ as it would be between two planes
at distance $L$
\begin{eqnarray}
\omega^2 - \omega_0^2 = \frac{b^2}I 2\pi R~P\left(L\right)~.
\label{IUPUIexpt}
\end{eqnarray}
$P$ is given by \eqref{CasimirRealP} and $\eta_\P$ will be discussed
in section~\ref{Lecture3}. Note that the separation $L$ between
bodies is measured separately through two-color interferometry, up
to a global offset adjusted in the data analysis
process~\cite{Decca2007prd,Decca2007epj}.

When compared with the theoretical prediction, this measurement
leads to unexpected conclusions~\cite{Decca2007prd,Decca2007epj}:
the measurements appear to agree with the predictions obtained from
the lossless plasma model $\gamma=0$ but to deviate significantly
from those deduced from the Drude model which accounts for
dissipation (see Fig.1 in~\cite{Decca2007prd}). These experiments
are performed in a range of distances $0.16\mu$m$<L<0.75\mu$m where
the difference between the predictions of the two models is small.
This entails that the problem of accuracy, that is also the control
of systematic errors, is a critical issue. However, the deviation of
experimental results from theoretical expectations (based on the
Drude model) is clearly larger than the statistical dispersion of
these results (bars on Fig.1 in~\cite{Decca2007prd}). More details
on statistical and systematic errors in this experiment can be found
in~\cite{Decca2007prd,Decca2007epj}.

Different conclusions are reached in a more recent experiment
performed in Yale~\cite{Sushkov2011}. This experiment aims at
measurements at larger distances $0.7\mu$m$<L<7\mu$m, where the
force is smaller while the thermal contribution and the effect of
dissipation are larger (see Fig.\ref{FigThermal}). The experimental
technique is based on a torsion balance and uses a much larger
sphere $R=156$mm, which allows for measurements of weaker pressures.
This experiment clearly sees the thermal effect and its results fit
the predictions drawn from the dissipative Drude model, after the
contribution of the electrostatic patch effect has been
subtracted~\cite{Sushkov2011}. These new results have to be
confirmed by further studies~\cite{Milton2011natphys}. The main
issue in this experiment is that the pressure due to electrostatic
patches is larger than that due to Casimir effect, so that a proper
modeling of this contribution is critical whereas the patch pattern
has not been characterized independently. This is in fact a more
general problem since the patch properties have not been measured in
other experiments either (more discussions below).

\subsection*{Discussion}

The conclusion at this point is that the Casimir effect is measured
with a good precision in several experiments, with a persisting
problem however in terms of accuracy. The results of the most
precise experiment, improved over a decade at IUPUI and confirmed
recently at UCR, appear to favor theoretical predictions obtained
with the lossless plasma model and to deviate from the predictions
obtained with the best motivated model, that is the
dissipative Drude model. The Yale experiment fits predictions drawn
from this Drude model, after the subtraction of a large
contribution of the electrostatic patch effect. For the IUPUI
experiment, the pressure difference goes up to $\sim$50mPa at the
smallest distances $\sim$160nm where the pressure itself is
$\sim$1000mPa, which entails that the accuracy is certainly not at
the 1\% level, as has been occasionally claimed.

\begin{figure}[tbh]
\includegraphics[width=0.5\textwidth]{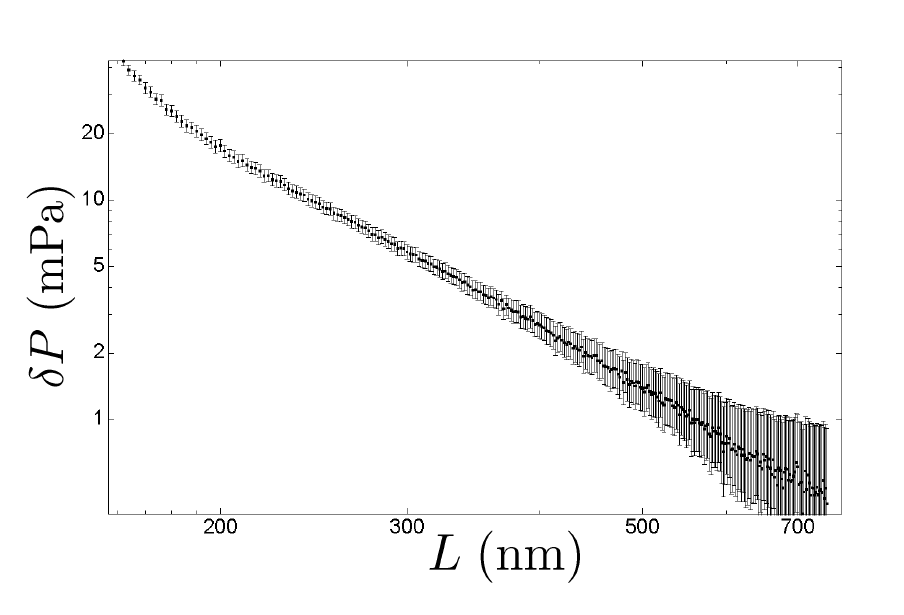}
\caption{Difference $\delta P=P_\mathrm{exp}-P_\mathrm{th}$ between
the experimental and theoretical values of the Casimir pressure as a
function of the distance $L$. Experimental values were kindly
provided by R.S. Decca and theoretical values calculated by R.O.
Behunin \emph{et al}~\cite{Behunin2012pra85}, with the Drude model
and at room temperature.} \label{FigDeviation}
\end{figure}

The difference $\delta P=P_\mathrm{exp}-P_\mathrm{th}$ between the
experimentally measured ($P_\mathrm{exp}$) and theoretically
predicted ($P_\mathrm{th}$) values of the Casimir pressure is drawn
on Fig.\ref{FigDeviation} as a function of the distance $L$.
Experimental values and error bars correspond to data kindly
provided by R.S. Decca~\cite{Decca2007prd,Decca2007epj,Decca2012pasi}.
Theoretical values were calculated by R.O. Behunin \emph{et al}
in~\cite{Behunin2012pra85}, using the optical data of gold
extrapolated at low frequencies to a Drude model, and at room
temperature. Systematic corrections were done in~\cite{Behunin2012pra85}
as in~\cite{Decca2007prd,Decca2007epj} and similar results
obtained. The discrepancy clearly appears on Fig.\ref{FigDeviation}
and it is of particular importance in the context of gravity tests
at sub-millimeter ranges~\cite{Adelberger2009,Antoniadis2011}. The
deviation seen on Fig.\ref{FigDeviation} does not look like a Yukawa
law, but it certainly looks like a combination of power laws~!

This discrepancy between theory and experiment may have various
origins, in particular artifacts in the experiments or inaccuracies
in the theoretical evaluations. They may also come from yet
unmastered systematic effects in the comparison between experimental
data and theoretical predictions. They could in principle be the
first hint of the existence of new forces beyond the standard model,
though such a strong statement should only be considered after a
cautious examination of the more mundane explanations associated in
particular with systematic effects.

The theoretical formula used to calculate the Casimir pressure
between real plates will be derived in section~\ref{Lecture3}. It
will reproduce the ideal Casimir expression at the limits of perfect
reflection and null temperature while being valid at any temperature
for any model of mirrors obeying well motivated physical
properties~\cite{Jaekel1991,Lambrecht2000}, including the case of
dissipative mirrors~\cite{Genet2003}. When the reflection amplitudes
are deduced from the Fresnel laws, and semi-infinite bulk mirrors
are characterized by a linear and local dielectric response
function, the results reproduce those of I.E. Dzyaloshinskii, E.M.
Lifshitz and L.P.
Pitaevskii~\cite{Lifshitz1956,Dzyaloshinskii1961,Lifshitz1980}. It
then remains to specify this dielectric function and its
low-frequency behavior. Here, it is worth emphasizing that the Drude
model, though being obviously much better motivated than the
lossless plasma model, is not a very accurate description of
conduction phenomena in real metals. More detailed descriptions can
be considered, which can for example be determined from microscopic
models of conduction in metals. Attempts in this direction and
discussions can be found for example
in~\cite{Pitaevskii2008prl,Dalvit2008prl,Svetovoy2008prl,%
Geyer2009prl,Pitaevskii2009prl,Decca2009prl,Dalvit2009prl}. To date,
they have been unable to explain the discrepancy.

A possible source of systematic error is the use of the Proximity
Force Approximation (PFA) in order to derive expressions for the
plane-sphere geometry from those known from the plane-plane
geometry. This approximation is expected to be valid at the limit
where the aspect ratio $x\equiv L/R$ goes to zero. Even in this
case, the accuracy of the PFA for a finite value of $x$ remains an
open question after the remarkable advances made recently on this
topic~\cite{Canaguier-Durand2012ijmpcs}, which will be described in
section \ref{Lecture3}. The question remains open, even though most
specialists would probably bet that the deviation from PFA is not
able to bridge the gap between experiment and theory. Other possible
sources of systematic error involve the effects of surface physics
on Casimir experiments. The problem of surface roughness has been
studied in a thorough
manner~\cite{Neto2005epl,Neto2005pra,Zwol2009,Broer2011,Broer2012,Broer2013}.

\subsection*{Electrostatic patches and contamination}

Electrostatic patches and contamination, already alluded to, are a
worrying source of such systematic effects, discussed in the sequel
of this section. Electrostatic patches have been known for
a long time to be a source of worries for a large number of
high precision measurements~\cite{Witteborn1967,Camp1991,%
Sandoghdar1996,Turchette2000,Deslauriers2006,Robertson2006,Epstein2007,%
Pollack2008,Dubessy2009,Carter2011,Everitt2011,Reasenberg2011,%
Hite2012,Hite2013}, and
in particular for Casimir experiments~\cite{Speake2003,Chumak2004,%
Kim2010pra,Kim2010jvst,Man2010} and short-range gravity
tests~\cite{Adelberger2009}. The patch effect is due to the fact
that the surface of a metallic plate cannot be an equipotential as
it is made of micro-crystallites with different work
functions. For clean metallic surfaces studied by the techniques of
surface physics, the resulting \emph{voltage roughness} is
correlated to the \emph{topography roughness} as well as to the
orientation of micro-crystallites~\cite{Gaillard2006}. For surfaces
exposed to air, the situation is changed due to the unavoidable
contamination by adsorbents, which spread out the
electrostatic patches, enlarge correlation lengths and reduce
voltage dispersions~\cite{Rossi1992}.

The pressure due to electrostatic patches between two planes can be
computed by solving the Poisson equation~\cite{Speake2003}. Its
evaluation depends on the spectra describing the correlations of the
patch voltages or, equivalently on the associated noise spectra
$C\left(k\right)$, with $k$ a patch wave-vector. In analysis of the
patch pressure devoted to Casimir experiments up to recently, the
spectrum was assumed to be flat between two \emph{sharp cutoffs} at
a minimum wave-vector $k_\min$ and a maximum one $k_\max$. Assuming
furthermore that $k_\min$ and $k_\max$ were given by the grain size
distribution measured with an Atomic Force Microscope (AFM), it was
concluded that the patch pressure was much smaller than the
discrepancy between experiment and
theory~\cite{Decca2005ap,Decca2007prd,Decca2007epj}.

A \emph{quasi-local} model has recently been proposed with the aim
of proposing a much better motivated representation of
patches~\cite{Behunin2012pra85}. The model is based on a
tessellation of the sample surface and a random assignment of the
voltage on each patch. It produces a smooth spectrum different from
the sharp-cutoff model used in previous analysis since there is now
contributions to the patch pressure coming from arbitrary low values
of $k$, even if the patch size distribution has an
upper bound. When the patch effect is estimated with the parameters
deduced from the grain size distribution as
in~\cite{Decca2005ap,Decca2007prd}, a much larger contribution of
patches is obtained. In fact, the calculated patch pressure is now
larger than the residuals between experimental data and theoretical
predictions, which means that patches could be a crucial systematic
effect for Casimir force measurements~\cite{Behunin2012pra85}.

As the computed patch pressure is model dependent, it seems natural
to try to find a model between the two cases presented above which
would reproduce at least qualitatively the residuals. By varying the
parameters of the quasi-local model, it was found
in~\cite{Behunin2012pra85} that the output of the model depended
mainly on two parameters, the size of largest patches
$\ell_\patch^\max$ and the rms voltage dispersion $V_\rms$ and that
a best-fit on these two parameters produced a qualitative agreement
between the residuals and the patch pressure. The best-fit values
for the parameters $\ell_\patch ^\max$ and $V_\rms$ are quite
different from those obtained by identifying patch and grain sizes.
With $\ell_\patch ^\max$ larger than the maximum grain size and
$V_\rms$ smaller than the rms voltage which would be associated with
random orientations of clean micro-crystallites, these values are
however compatible with a contamination of metallic surfaces, which
had to be expected anyway.

It follows that the difference between IUPUI experimental data and
theoretical predictions can be fitted at least qualitatively by a
simple model for electrostatic patches. This conclusion is however
only the result of a fit, with the parameters of the patch model not
measured independently. In order to reach a firm conclusion, the
patch spectrum has to be measured independently, by using the
dedicated technique of Kelvin probe force microscopy (KPFM) which is
able to achieve the necessary size and voltage
resolutions~\cite{Liscio2008jpc,Liscio2011acr}. When these
characteristics are available, the contribution of the patches to
the Casimir measurements can be evaluated and unambiguously
subtracted when comparing theory and experiments.

Preliminary results of such characterizations have recently been
published~\cite{Behunin2014}. Note that the evaluation of force was
done for the plane-sphere geometry~\cite{Behunin2012pra86}.

\section{A simple derivation of the Casimir effect in one dimension}
\label{Lecture2}

The present section \ref{Lecture2} contains a derivation of the
Casimir effect in a model of scalar fields propagating along the two
directions on a 1-dimensional line. Within this simple model which
lays the basis for more complicated calculations to appear in the
next section, we introduce the \emph{Quantum Optics} approach to the
Casimir effect. The approach is based on the scattering of vacuum
field fluctuations obtained in the ground state of the associated
\emph{Quantum Field Theory}. Each mirror is described by a
scattering operator~\cite{Lippmann1950,GellMann1953} which is
reduced here to a $2\times2$ matrix containing reflection and
transmission amplitudes. Two mirrors form a Fabry-Perot cavity with
all field transformations deduced from the two elementary scattering
matrices. The Casimir force then results from the difference of
radiation pressures exerted onto the inner and outer sides of the
mirrors by the vacuum field fluctuations. Equivalently, the Casimir
free energy can be written as the shift of field energy due to the
presence of the Fabry-Perot cavity~\cite{Schwinger1975,Plunien1986}.
The formula obtained in this manner is valid and regular at thermal
equilibrium at any temperature and for any optical model of mirrors
obeying causality and high frequency transparency properties.

The radiation pressure interpretation of the Casimir force was
presented for perfect mirrors in~\cite{Milonni1988} and extended to
the case of real mirrors~\cite{Jaekel1991}. The calculations were
then systematically expanded in particular for applications to the
problem of the \emph{Dynamical Casimir
effect}~\cite{Jaekel1992qo,Jaekel1992jp,Jaekel1992pla,Jaekel1993jp1,%
Jaekel1993jp2,Jaekel1993pla,Lambrecht1996prl,Lambrecht1998epjd}. It
has also served as a basis for the Scattering Formalism for the
static Casimir effect~\cite{Genet2003,Lambrecht2006} of which we
will give a pedagogical presentation in the following.

\subsection{Quantum field theory on the one-dimensional line}

We consider here quantum field theory on the one-dimensional line,
that is also quantum field theory in two-dimensional space-time (one
time coordinate $t$, one space coordinate $x$). The field
propagation is thus described by the d'Alembert's wave equation,
originally written for the propagation of transverse vibrations of a
string, and which also describes many wave phenomena such as
electrical propagation in a transmission line, acoustic wave and so
on.

\subsubsection*{Propagation equation on the one-dimensional line} We
write it here for a single vibration described by the scalar
potential $\Phi(x,t)$
\begin{equation}
\frac{\partial^2\Phi}{\partial t^2} - c^2
\frac{\partial^2\Phi}{\partial x^2} =0 ~. \label{d'AlembertEquation}
\end{equation}
The general solution to this equation is given by the d'Alembert's
formula, that is the superposition of rightward ($\varphi^+$) and
leftward ($\varphi^-$) traveling waves propagating at the velocity
$c$ in opposite directions along the $x$-axis
\begin{equation}
\Phi(t,x)=\varphi^+\left(u_+\right) + \varphi^-\left(u_-\right)~,
\label{d'AlembertFormula}
\end{equation}
where $u_\pm$ are called today the light cone variables
\begin{equation}
u_+\equiv t-\frac {x}c\quad,\quad u_-\equiv t+\frac {x}c ~.
\label{lightconevariables}
\end{equation}

In this simplest version of field theory, there is one normal mode
for each frequency $\omega \in [0,\infty]$ and each propagation
direction $\eta=\pm1$. The standard methods of quantum field
theory~\cite{Itzykson1985,Cohen-Tannoudji1992} then allow one to
write the rightward ($\varphi^+$) and leftward ($\varphi^-$)
traveling waves as Fourier decompositions over canonical mode
operators
\begin{eqnarray}
\varphi^\eta\left(u\right) = {{\int}}_0^\infty \,\frac{\dd
\omega}{2\pi} \sqrt{\frac\hbar{2\omega}} \left(a_{\omega,\eta}
e^{-\imag\omega u} +a_{\omega,\eta}^\dag e^{\imag\omega u} \right)~.
\label{varphiFourier}
\end{eqnarray}
Annihilation and creation operators $a_{\omega,\eta}$ and
$a_{\omega,\eta}^\dag$ correspond respectively to positive and
negative frequencies in the decompositions \eqref{varphiFourier}.
Note that the fields in space-time $ \varphi^\eta(u)$ are
real-valued, so that annihilation and creation operators are
hermitian conjugate of each other. They obey the following canonical
commutation relations
\begin{eqnarray}
\label{commutationrelations1D}
&\left[ \, a_{\omega,\eta} \,,\,
a_{\omega^\prime,\eta^\prime}^\dag \,\right]=
2\pi \delta \left(\omega-\omega^\prime\right ) \delta_{\eta,\eta^\prime}~,&\\
&\left[ \, a_{\omega,\eta} \,,\, a_{\omega^\prime,\eta^\prime}
\,\right]= \left[ \, a_{\omega,\eta}^\dag \,,\,
a_{\omega^\prime,\eta^\prime} ^\dag\,\right]= 0&~. \nonumber
\end{eqnarray}

The Hamiltonian $\calH$ for the d'Alembert's wave equation
\eqref{d'AlembertEquation} is the integral over space of the energy
density $e(t,x)$
\begin{eqnarray}
\calH = \int \dd x ~ e(t,x) \;,\; e \equiv  \frac12 \left(
\frac{\partial\Phi}{\partial t} \right)^2 +  \frac{c^2}2 \left(
\frac{\partial\Phi}{\partial x} \right)^2.
\end{eqnarray}
Using the d'Alembert's formula \eqref{d'AlembertFormula} for fields,
one derives another d'Alembert's formula describing the general
energy density as the superposition of rightward and leftward
traveling flows
\begin{eqnarray}
e(t,x) = e^+\left(u_+\right)+ e^-\left(u_-\right) \;,\;
e^\eta\left(u_\eta\right) =
\left(\frac{\partial\varphi^\eta}{\partial u_\eta} \right)^2.
\label{energydensities}
\end{eqnarray}

\subsubsection*{Vacuum and thermal fluctuations}

Vacuum $\left\vert\, \mathrm{vac} \,\right\rangle$ is the
fundamental state of quantum field with an infinite number of
modes each containing no photons. This means that all annihilation
operators vanish in this state while creation operators have their
action determined by the commutation relations
\eqref{commutationrelations1D}.

In the following, we use the correlation functions of vacuum fields
which are deduced from these elementary properties
\begin{eqnarray}
& \left\langle \, a_{\omega,\eta}  \,\right\rangle_\mathrm{vac}  =
\left\langle \, a_{\omega,\eta}^\dag  \,\right\rangle_\mathrm{vac} =
0 ~,& \nonumber \\
& \left\langle \, a_{\omega,\eta} a_{\omega^\prime,\eta^\prime}^\dag
\,\right\rangle_\mathrm{vac}  = 2\pi \delta
\left(\omega-\omega^\prime\right ) \delta_{\eta,\eta^\prime} ~,&
\nonumber \\ & \left\langle \, a_{\omega,\eta}^\dag
a_{\omega^\prime,\eta^\prime} \,\right\rangle_\mathrm{vac}  = 0 ~,&
\nonumber \\
& \left\langle \, a_{\omega,\eta} a_{\omega^\prime,\eta^\prime}
\,\right\rangle_\mathrm{vac} = \left\langle \, a_{\omega,\eta}^\dag
a_{\omega^\prime,\eta^\prime}^\dag \,\right\rangle_\mathrm{vac} = 0
~,& \label{correlationVacuum}
\end{eqnarray}
where $\left\langle \, \ldots \, \right\rangle_\mathrm{vac} \equiv
\left\langle \, \mathrm{vac} \, \right\vert \, \ldots \, \left\vert
\, \mathrm{vac}  \, \right\rangle$. Higher-order correlation
functions are deduced from the fact that vacuum fields may be dealt
with as Gaussian random variables.

In a state at thermal equilibrium at temperature $T$, the first and
last lines in \eqref{correlationVacuum} are unchanged whereas the
second and third lines are changed to
\begin{eqnarray}
& \left\langle \, a_{\omega,\eta} a_{\omega^\prime,\eta^\prime}^\dag
\,\right\rangle_\mathrm{therm}  = 2\pi \delta
\left(\omega-\omega^\prime\right )
\delta_{\eta,\eta^\prime} \left(1+\ovn\right) ~,& \nonumber \\
&\left\langle \, a_{\omega,\eta}^\dag a_{\omega^\prime,\eta^\prime}
\,\right\rangle_\mathrm{therm}  = 2\pi \delta
\left(\omega-\omega^\prime\right ) \delta_{\eta,\eta^\prime} ~\ovn
~,& \label{correlationThermal}
\end{eqnarray}
where $\left\langle \, \ldots \, \right\rangle_\mathrm{therm} \equiv
\left\langle \, \mathrm{therm} \, \right\vert \, \ldots \,
\left\vert \, \mathrm{therm}  \, \right\rangle$ while $\ovn$ is the
mean photon number in Planck's law \eqref{Planck1911}. Note that
\eqref{correlationThermal} is reduced to \eqref{correlationVacuum}
when $T\to0$. Note also that all field commutators are unchanged,
which is consistent with the fact that they are directly connected
to the propagators.

Using  the Fourier decompositions \eqref{varphiFourier} of the
fields and the correlation functions \eqref{correlationThermal}, we
deduce that the mean values of energies densities
\eqref{energydensities} have the following spectral decompositions
in the general case ($T\neq0$)
\begin{eqnarray}
\left\langle \, e^\eta\left(u\right) \,\right\rangle_\mathrm{therm}
&=& {{\int}}_0^\infty \,\frac{\mathrm{d}\omega}{2\pi}
{{\int}}_0^\infty \,\frac{\mathrm{d}\omega^\prime}{2\pi}
 \sqrt{\frac{\hbar\omega}2}
 \sqrt{\frac{\hbar\omega^\prime}2} \nonumber \\
&&\times \left\langle \, a_{\omega,\eta}
a_{\omega^\prime,\eta^\prime}^\dag + a_{\omega,\eta}^\dag
a_{\omega^\prime,\eta^\prime} \,
\right\rangle_\mathrm{therm} \nonumber \\
&=& {{\int}}_0^\infty \,\frac{\mathrm{d}\omega}{2\pi}
{\frac{\hbar\omega}2} \left(1+2\ovn\right) ~.
\label{meanenergiesThermal}
\end{eqnarray}
This is the modern expression of the second Planck's
law~\cite{Planck1911} for the energy per mode given as the sum of
vacuum and thermal contributions (compare with \eqref{Planck1911}).
The limit of zero temperature corresponds to $\ovn=0$ in this
expression.

\subsection{One mirror on a 1d line}

\begin{figure}[tbh]
\includegraphics[width=0.35\textwidth]{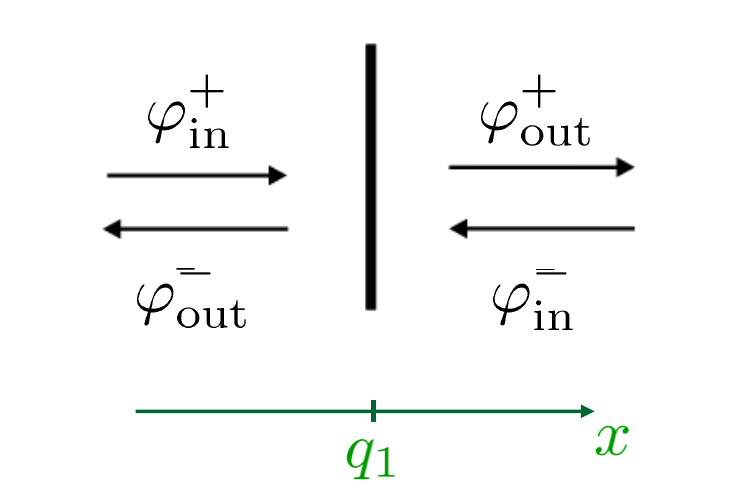}
\caption{Schematic representation of a mirror at position $q_1$ on a
1d line~: this mirror is a point-like scatterer which couples
rightward and leftward propagating waves.} \label{FigMirror1d}
\end{figure}

We now consider the situation where one mirror is placed into vacuum
at a position $q_1$ (see Fig.\ref{FigMirror1d}).

\subsubsection*{Scattering by one mirror on a 1d line}
The general solution is in this case a generalized d'Alembert's
formula with different expressions for fields on the lefthand
($\Phi_\L $) and righthand ($\Phi_\R $) sides of the mirror
\begin{eqnarray}
&\Phi(t,x) = \Phi_\L (t,x) + \Phi_\R (t,x)~,& \\
&\Phi_\L (t,x) = \varphi^+_\iin  (u_+) +
\varphi^-_\out  (u_-) \quad,\quad x<q_1~,& \nonumber\\
&\Phi_\R (t,x) = \varphi^+_\out  (u_+) + \varphi^-_\iin (u_-)
\quad,\quad q_1<x  ~.& \nonumber
\end{eqnarray}
The symbol $+$ and $-$ represent as previously rightward and
leftward propagation while the symbols in and out correspond to
incoming and outgoing traveling waves. These waves are coupled by
scattering on the mirror.

For the limiting case of perfectly reflecting mirrors, the field
$\Phi$ vanishes at the left and right sides of the mirrors. Output
fields are then easily deduced from input ones as $\varphi^\pm_\out
\left(t,q_1\right) = - \varphi^\mp_\iin \left(t,q_1\right)$. In
Fourier space, this is written as a scattering process with
reflection amplitudes having a unit modulus and a phase determined
by the position of the mirror, $ \varphi^\pm _\out [\omega] = -
e^{\mp 2ikq_1} \varphi^\mp _\iin [\omega]$ with $k\equiv\omega/c$.
As the scattering process preserves the frequency for a motionless
mirror, these equations could as well have been written for
annihilation ($\omega>0$) and creation ($\omega<0$) operators.

\subsubsection*{Real mirrors on a 1d line}
Real mirrors cannot be perfectly reflecting at all frequencies. They
are described by a more general scattering matrix containing
transmission as well as reflection amplitudes (with
$k\equiv\omega/c$)
\begin{eqnarray}
&&\left(
\begin{array}{c}
\varphi^+ _\out [\omega] \\
\varphi^-_\out [\omega]
\end{array}
\right)  = S_1 [\omega] \left(
\begin{array}{c}
\varphi^+ _\iin [\omega] \\
\varphi^- _\iin [\omega]
\end{array}
\right) ~, \nonumber \\
&&S_1 [\omega] = \left(
\begin{array}{cc}
t_1 [\omega] & r_1[\omega] e^{-2ikq_1} \\
r_1[\omega] e^{2ikq_1} & t_1[\omega]
\end{array}
\right)~. \label{RealMirror}
\end{eqnarray}
The scattering amplitudes $r_1$ and $t_1$ have been defined for a
mirror located at $x=0$ and the general case then obtained by
introducing the phases $e^{\mp 2ikq_1}$ determined by the position
of the mirror. We have considered the particular case of a
symmetrical scattering matrix for a mirror located at $x=0$. A more
general treatment would not change any important result in the
following.

The scattering matrix preserves frequency since energy is conserved
for a stationary scattering but it depends on frequency as a
consequence of fundamental physical properties~\cite{Jaekel1991}.
The scattering process considered here obeys the following
properties (the unitarity assumption, valid only for lossless
mirrors, will be released later on)
\begin{enumerate}
\item{Fields are real in the time domain, so that
$S[\omega]^{\dag}=S[-\omega]$, that is also
$t_1[\omega]^\ast=t_1[-\omega]$ and
$r_1[\omega]^\ast=r_1[-\omega]$;}
\item{The scattering process obeys causality, so that the
amplitudes can be prolongated as analytical functions in the upper
half of the complex plane (more details below);}
\item{The scattering process obeys unitarity, so that
$S[\omega] S[\omega]^{\dag}=\mathcal{I}$, that is also
$|r_1[\omega]|^2 + |t_1[\omega]|^2 =1$ and
$t_1[\omega]r_1[\omega]^\ast + r_1[\omega]t_1[\omega]^\ast =0$; }
\item{Reflection tends to vanish at the high-frequency limit
$\displaystyle{\lim_{\omega \rightarrow \infty} S_1[\omega]}
\rightarrow \mathcal{I}$, so that $\displaystyle{\lim_{\omega
\rightarrow \infty} t_1[\omega]} \rightarrow 1$ and
$\displaystyle{\lim_{\omega \rightarrow \infty} r_1[\omega]}
\rightarrow 0$.}
\end{enumerate}

These general properties may be illustrated with an example, which
corresponds in particular to a transmission line with a localized
impedance mismatch~\cite{Jaekel1993pla}. For this simple example,
the d'Alembert's wave equation \eqref{d'AlembertEquation} is changed
to
\begin{eqnarray}
\frac{\partial^2\Phi}{\partial t^2}
-c^2\frac{\partial^2\Phi}{\partial x^2} +2c\Omega \delta(x-q_1)
\Phi(t,q) =0~,
\end{eqnarray}
and the solution obtained as \eqref{RealMirror} with
\begin{eqnarray}
r_1[\omega] = \frac{\Omega} {i\omega-\Omega}\quad,\quad t_1[\omega]
= \frac{i\omega} {i\omega-\Omega}~. \label{ModelMirror}
\end{eqnarray}
The general properties (1-4) enumerated in the preceding paragraph
can easily be checked out. The parameter $\Omega$ appears as the
physical cutoff above which reflection tends to vanish. The limit of
perfect reflection can be defined by the case of large values of
$\Omega$. It has been shown that this definition allows one to
escape the difficulties encountered when studying perfectly
reflecting mirrors~\cite{Jaekel1992pla,Jaekel1993jp1}.

\subsubsection*{Force on one mirror on the 1d line}
\begin{figure}[tbh]
\includegraphics[width=0.35\textwidth]{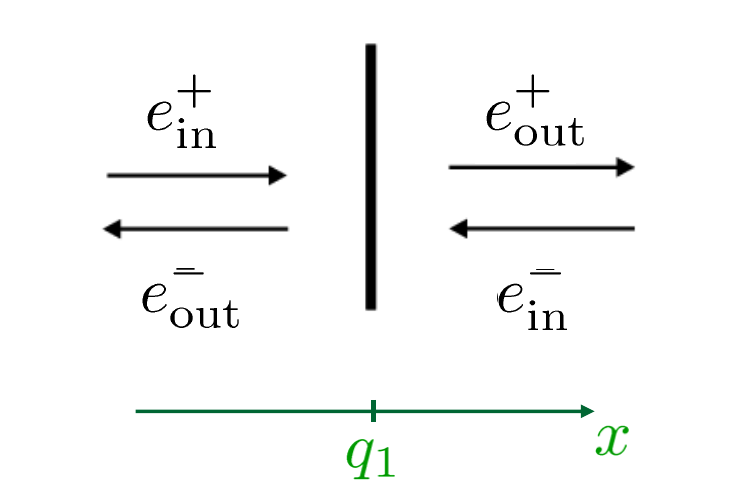}
\caption{Schematic representation of energy flows on a 1d line due
to scattering on a mirror at position $q_1$ (see
Fig.\ref{FigMirror1d}).} \label{FigEnergy1d}
\end{figure}

We come now to the evaluation of the force acting on the mirror
represented on figure \ref{FigMirror1d}. To this aim, we first study
the energy flows in the same situation, as sketched on figure
\ref{FigEnergy1d}.

The general solution is now a d'Alembert's formula
\eqref{energydensities} for energy densities with different
expressions on the lefthand ($e_\L $) and righthand ($e_\R $) sides
of the mirror
\begin{eqnarray}
\label{energydensities1mirror}
&e(t,x) = e_\L (t,x) + e_\R (t,x)~,& \\
&e_\L (t,x) = e^+_\iin  (u_+) +
e^-_\out  (u_-) \quad,\quad x<q_1~,& \nonumber\\
&e_\R (t,x) = e^+_\out  (u_+) + e^-_\iin  (u_-) \quad,\quad q_1<x
~.& \nonumber
\end{eqnarray}
Similar expressions can be written for the momentum densities by
just putting a sign $\eta$ in front of energy densities $e^\eta$.
The force on the mirror is then deduced from a momentum balance upon
scattering and it is found to be proportional to the difference of
the energy densities $e_\L (t,q)$ and $e_\R (t,q)$ on the lefthand
and righthand sides of the mirror~\cite{Jaekel1992qo}.

In the limiting case of perfectly reflecting mirrors, the field
$\Phi$ vanishes at the left and right sides of the mirrors. Output
fields are then easily deduced from input ones as $e^\pm_\out
\left(t,q_1\right) = e^\mp_\iin  \left(t,q_1\right)$, and the force
thus obtained as $F(t) = 2 \left(e_\iin ^+(t,q)-e_\iin
^-(t,q)\right)/c$. The mean values $\left\langle e_\iin ^\pm(t,q)
\right\rangle_\mathrm{therm}$ of leftward and rightward energy
densities are infinite (see \eqref{meanenergiesThermal}). But these
mean values are also equal for leftward and rightward densities, so
that the force on the mirror is zero $\left\langle F
\right\rangle_\mathrm{therm}=0$.

This result crucially depends on the fact that we have evaluated the
mean force on a mirror at rest. There exist non vanishing
fluctuations of force on the mirror, as well as a non null mean
force for a moving mirror~\cite{Jaekel1992qo}.

\subsubsection*{Force on one real mirror}
In the general case of a non perfect mirror, the same result is
proven by the following reasoning. First, the force is deduced from
the momentum balance upon scattering and the expressions
\eqref{energydensities1mirror} of energy densities
\begin{eqnarray}
\label{force1mirror} F(t) &=& \frac{e_\L (t,q)-e_\R (t,q)}c
\\ &=& \frac{ e^+_\iin (t,q) + e^-_\out  (t,q) - e^+_\out
(t,q) - e^-_\iin (t,q) }c~. \nonumber
\end{eqnarray}
Then, the mean energy densities are calculated by using the
expressions \eqref{energydensities} of energy densities and the
description \eqref{RealMirror} of the scattering process. From the
unitarity of the $S-$matrix, one deduces that mean radiation
pressures are still equal on the two sides (equation written at
thermal equilibrium at temperature $T$)
\begin{eqnarray}
\left\langle \, e_\out ^+ \,\right\rangle &=& \left\langle \, e_\out
^- \,\right\rangle =\left\langle \, e_\iin ^+ \,\right\rangle
=\left\langle \, e_\iin ^- \,\right\rangle \nonumber \\
&=& {{\int}}_0^\infty \,\frac{\mathrm{d}\omega}{2\pi} \,\hbar\omega
\left(\frac12+\ovn\right)~, \label{meanenergydensities1mirror}
\end{eqnarray}
so that the mean value of the force still vanishes
\begin{eqnarray}
\left\langle \, F \,\right\rangle = 0~.
\end{eqnarray}

\subsection{Two mirrors on the 1d line}

\begin{figure}[tbh]
\includegraphics[width=0.4\textwidth]{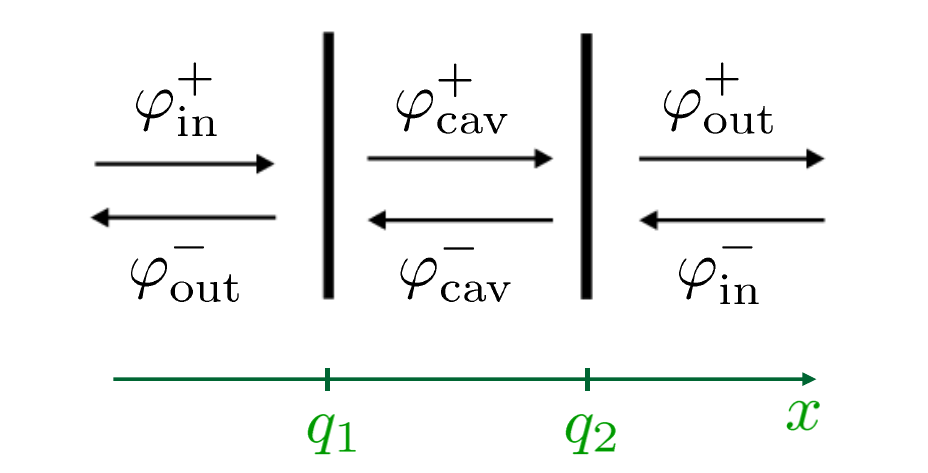}
\caption{Schematic representation of the scattering of fields by two
mirrors at positions $q_1$ and $q_2$ on a 1d line.}
\label{FigCavity1d}
\end{figure}

The situation changes fundamentally as soon as we consider that
there are two mirrors present on the 1d line, which form a
Fabry-Perot cavity. There is now a difference between the inner and
outer sides of the mirrors, as sketched on Figure \ref{FigCavity1d}.

\subsubsection*{Scattering by a cavity on a 1d line}
The spatial positions of the two mirrors are denoted $q_1$ and $q_2$
with the length of the cavity $L\equiv q_2-q_1>0$. Each mirror
couples rightward and leftward traveling waves. In contrast with the
case of one mirror, the fields undergo multiple scattering and there
appears an intra-cavity region. The general solution for the fields
is now written with different expressions for fields on the lefthand
($\Phi_\L $) and righthand ($\Phi_\R $) sides of the cavity and
those ($\Phi_\C $) within the cavity
\begin{eqnarray}
&\Phi(t,x) = \Phi_\L (t,x) + \Phi_\C (t,x) + \Phi_\R (t,x)~,& \\
&\Phi_\L (t,x) = \varphi^+_\iin  (u_+) +
\varphi^-_\out  (u_-) \;,\; x<q_1~,& \nonumber\\
&\Phi_\C (t,x) = \varphi^+_\cav  (u_+) +
\varphi^-_\cav  (u_-) \;,\; q_1<x<q_2~,& \nonumber\\
&\Phi_\R (t,x) = \varphi^+_\out  (u_+) + \varphi^-_\iin (u_-) \;,\;
q_2<x  ~.& \nonumber
\end{eqnarray}

The scattering effect of the cavity has now to be described by two
matrices, instead of one in the one mirror case, a global scattering
matrix $S$ which gives the output fields in terms of the input ones,
and a resonance matrix $R$ which gives the intra-cavity fields
\begin{eqnarray}
&&\left(
\begin{array}{c}
\varphi^+ _{\out} \\
\varphi^-_{\out}
\end{array}
\right)  = S \left(
\begin{array}{c}
\varphi^+ _{\iin}  \\
\varphi^- _{\iin}
\end{array}
\right) ~, \nonumber \\ &&\left(
\begin{array}{c}
\varphi^+ _{\cav} \\
\varphi^-_{\cav}
\end{array}
\right)  = R \left(
\begin{array}{c}
\varphi^+ _{\iin}  \\
\varphi^- _{\iin}
\end{array}
\right)~. \label{SandRmatrices}
\end{eqnarray}

\subsubsection*{Scattering and resonance matrices}
The global $S-$matrix can be evaluated from the elementary matrices
$S_1$ and $S_2$ associated with the two mirrors and the free
propagation phase-shifts. The results can be written as follows (all
amplitudes depend on $\omega\equiv ck$)
\begin{eqnarray}
S = \frac 1d \left(
\begin{array}{cc}
t_1 t_2 & d r_2 e^{-ikL}+
t_2^2 r_1 e^{ikL} \\
d r_1 e^{-ikL}+ t_1^2 r_2 e^{ikL} & t_1 t_2
\end{array}
\right) ~. \nonumber\\ \label{globalSmatrix}
\end{eqnarray}
The denominator $d$ is an important function, with its zeros
corresponding to the resonances of the cavity,
\begin{eqnarray}
d[\omega] = 1-r[\omega] e^{2ikL} \quad,\quad r[\omega] \equiv
r_1[\omega] r_2[\omega]~. \label{denominator}
\end{eqnarray}
For the problem under consideration, the global $S-$matrix obeys the
same properties as the elementary matrices $S_1$ and $S_2$. In
particular, it is unitary
\begin{eqnarray}
S[\omega] S[\omega]^\dag = \mathcal{I}~. \label{globalunitarity}
\end{eqnarray}

The resonance matrix can be deduced from the elementary matrices
$S_1$ and $S_2$ associated with the two mirrors and the free
propagation phase-shifts. The results are obtained as (amplitudes
depend on $\omega$ and expressions are simplified by assuming $q_2 =
- q_1 = L/2$)
\begin{eqnarray}
R = \frac 1d \left(
\begin{array}{cc}
t_1  &
t_2^2 r_1 e^{ikL} \\
t_1^2 r_2 e^{ikL} & t_2
\end{array}
\right)~. \label{Rmatrix}
\end{eqnarray}
The resonance matrix shares some of the properties listed above
with the $S-$matrix but it is not unitary. It turns out that $R
R^\dag$ can be written~\cite{Jaekel1992jp}
\begin{eqnarray}
&&R[\omega] R[\omega]^\dag = \mathcal{I} + Q[\omega] +
Q[\omega]^\dag~,\nonumber\\ &&Q = \frac 1d \left(
\begin{array}{cc}
r_1 r_2 e^{2ikL} & r_1 e^{ikL} \\
r_2 e^{ikL} & r_1 r_2 e^{2ikL}
\end{array}
\right)~. \label{Qmatrix}
\end{eqnarray}

\subsection{Casimir force on the 1d line}

We now evaluate the energy densities and the forces in the case of
two mirrors on the 1d line.

\subsubsection*{Energy densities}
The general solution for the energy densities is now written as in
\eqref{energydensities1mirror}, considering the configuration
sketched on Figure \ref{FigCavity1d} for the cavity.
\begin{eqnarray}
&e(t,x) = e_\L (t,x) + e_\C (t,x) + e_\R (t,x)~,& \\
&e_\L (t,x) = e^+_\iin  (u_+) +
e^-_\out  (u_-) \;,\; x<q_1~,& \nonumber\\
&e_\C (t,x) = e^+_\cav  (u_+) +
e^-_\cav  (u_-) \;,\; q_1<x<q_2~,& \nonumber\\
&e_\R (t,x) = e^+_\out  (u_+) + e^-_\iin  (u_-) \;,\; q_2<x ~.&
\nonumber
\end{eqnarray}
We deduce expressions for the forces acting on each mirror 1 or 2,
obtained as differences of the radiation pressures on the lefthand
and righthand sides of the mirror (compare with
\eqref{force1mirror})
\begin{eqnarray}
&&F_\mathrm{1}(t) = e_\L  (t,q_1) - e_\C  (t,q_1) ~, \nonumber\\
&&F_\mathrm{2}(t) = e_\C  (t,q_2)  - e_\R  (t,q_2)~.
\end{eqnarray}

The energy densities on the outer sides of the cavity have the same
expressions as in the one-mirror case, provided that the scattering
matrix \eqref{globalSmatrix} of the cavity is used. As this global
$S-$matrix is unitary (see \eqref{globalunitarity}), one deduces
that the mean energy densities on the outer sides are still given by
equations \eqref{meanenergydensities1mirror}. Hence, they are
infinite but equal on the two sides of the cavity so that the mean
value of the global force on the cavity vanishes
\begin{eqnarray}
\left\langle \, F_1 + F_2 \,\right\rangle = 0~. \label{globalforce}
\end{eqnarray}

\subsubsection*{Intra-cavity energy densities}
The situation is different for the energy densities on the
intra-cavity sides of the mirrors, since their calculation is now
determined by the properties of the resonance matrix. Using the
property \eqref{Qmatrix}, one deduces that the mean values of these
energy densities is given by~\cite{Jaekel1992jp}
\begin{eqnarray}
\left\langle \, e_\cav ^+ \,\right\rangle =\left\langle \, e_\cav ^-
\,\right\rangle = {{\int}}_0^\infty \,\frac{\mathrm{d}\omega}{2\pi}
\,\hbar\omega \left(\frac12+\ovn\right)~g[\omega]~,
\label{meanenergydensitiescavity}
\end{eqnarray}
where $g[\omega]$ is the common value of the diagonal elements in
the matrix $R R^\dag$
\begin{eqnarray}
&&g[\omega] \equiv 1 + f[\omega] + f[\omega]^\ast = \frac{ 1 - \vert
r \vert^2}{ \vert 1 - r \, e^{2ikL} \vert^2} ~, \nonumber\\
&&f[\omega] \equiv \frac {r e^{2ikL}}{1-r e^{2ikL}}  ~.
\label{gandf}
\end{eqnarray}

The real function $g[\omega]$ represents the modification of the
energy density inside the cavity with respect to that outside the
cavity, which is in fact identical for input and output fields. The
same result would have been obtained for a classical calculation
with an input field at frequency $\omega$. Here this function
$g[\omega]$ describes the change of energy densities for vacuum as
well as thermal fluctuations. Its relation \eqref{gandf} to the
\emph{closed loop function} $f[\omega]$ will be used in the
following to transform the expression of the Casimir force.

\subsubsection*{Casimir force as a result of radiation pressures}
Collecting these results, one deduces the following expression of
the Casimir force, defined as the mean force on the righthand mirror
or the opposite of that on the lefthand mirror (see
\eqref{globalforce}),
\begin{eqnarray}
F &\equiv& \left\langle \, F_2 \,\right\rangle = -\left\langle \,
F_1 \,\right\rangle \nonumber \\ &=& {{\int}}_0^\infty
\,\frac{\mathrm{d}\omega}{2\pi c} \, \hbar\omega
\left(1+2\ovn\right) \left( g[\omega] -1 \right) ~.
\label{casimirforce1dg}
\end{eqnarray}
This expression is obtained as the difference of radiation pressures
\eqref{meanenergydensitiescavity} and
\eqref{meanenergydensities1mirror} on the inner and outer sides of
the mirrors.

Resonant frequencies correspond to an increase of energy in the
cavity ($g>1$) and they produce repulsive contributions to the
Casimir force. In contrast, frequencies out of resonance correspond
to a decrease of energy in the cavity ($g<1$) and they produce
attractive contributions to the Casimir force. The net force is the
integral of these contributions over all modes. This interpretation
of the Casimir force as a result of radiation pressures of vacuum
and thermal fluctuations produces a final expression which is finite
for any properly defined model of mirrors~\cite{Jaekel1991}. This is
seen more easily by using causality properties to rewrite
\eqref{casimirforce1dg} as an integral over imaginary frequencies.
Let us stress at this point that this rewriting is just a
mathematical transformation which does not affect the physical
content of \eqref{casimirforce1dg}. The rewriting will however spoil
its direct intelligibility as imaginary frequencies do not
correspond to physical modes.

\subsubsection*{Casimir force as an integral over imaginary frequencies}
One now rewrites the Casimir force as an integral over imaginary
frequencies by using the causality properties of the scattering
amplitudes. We give here a simplified description (a more general
derivation can be found in~\cite{Guerout2014}).

We first write the Casimir force \eqref{casimirforce1dg} as the real
part of a complex integral $F_r$ defined over the positive part
$\mathbb{R}^+$ of the real axis
\begin{eqnarray}
F &=& F_r + F_r^\ast ~,\nonumber\\ F_r &\equiv& {{\int}}_0^\infty
\,\frac{\mathrm{d}z}{2\pi c} ~ \hbar z ~f[z]~\coth\frac{\hbar
z}{2k_\B T} ~. \label{casimirforce1df}
\end{eqnarray}
We have used the relation \eqref{gandf} between $g$ and $f$,
substituted the frequency $\omega$ by a complex variable $z$ running
over $\mathbb{R}^+$. We have also replaced $1+2\ovn[\omega]$ by its
explicit form \eqref{Planckmodern}. Now the closed loop function
$f[z]$ is defined from causal reflection amplitudes and propagation
phases. Considered as a function of $z\in\mathbb{C}$, it has poles
in the lower half part of the complex plane which correspond to
resonances of the Fabry-Perot cavity, while it is analytical in the
upper half part of the complex plane $\Im z\ge0$. Meanwhile the
function $\displaystyle{\coth \left(\hbar z/2k_\B T\right)}$ is
analytical in the right half part of the complex plane $\Re z>0$ but
has poles at the Matsubara frequencies which are regularly spaced on
the imaginary axis~\cite{Matsubara1955}
\begin{eqnarray}
z_n= \imag\xi_n \quad,\quad \xi_n = n \xi_1 \quad,\quad \xi_1 =
\frac{2\pi k_\B T}{\hbar}~. \label{Matsubarapoles}
\end{eqnarray}

It follows that the integral $F_r$ can be transformed by using
Cauchy's theorem. Precisely, we apply the Cauchy's theorem to the
integral of the integrand appearing in \eqref{casimirforce1df} over
a closed contour consisting of $\mathbb{R}^+$, the positive part of
the imaginary axis shifted by a small positive real number and a
quarter of a circle with a very large radius. This last part
vanishes as a consequence of the high frequency transparency of the
mirrors. The integral over the whole contour also vanishes since the
integrand is an analytical function in the domain enclosed by the
contour. As a consequence, the integral $F_r$ may be written under
the equivalent form $F_i$ ($\varepsilon \to 0^+$)
\begin{eqnarray}
F_r = F_i = {{\int}}_{\varepsilon}^{\varepsilon+\imag\infty}
\,\frac{\mathrm{d}z}{2\pi c} ~ \hbar z ~f[z]~\coth\frac{\hbar
z}{2k_\B T} ~. \label{casimirforce1dfi}
\end{eqnarray}
The same transformation is then performed for $F_r^\ast$ which is
the integral of the same function over the negative part
$\mathbb{R}^-$ of the real axis, run from $-\infty$ to 0. $F_r^\ast$
is equal to $F_i^\ast$, the integral of the same integrand over the
positive part of the imaginary axis shifted by a small negative real
number $-\varepsilon$ and run from $-\varepsilon-\imag\infty$ to
$-\varepsilon$.

In the end, the Casimir force \eqref{casimirforce1df} is the
integral over a contour which encircles the imaginary axis, and it
is thus found to be a discrete sum of the values of the function $z
f[z]$ at the Matsubara poles $z_n$
\begin{eqnarray}
&&F=-2 k_\B T \sum_n^\prime \frac{\kappa_n r[\imag c\kappa_n]
e^{-2\kappa_n L}} {1-r[\imag c\kappa_n] e^{-2\kappa_n L}} ~,
\nonumber\\ &&\kappa_n \equiv \frac{\xi_n}c =n \kappa_1 \quad,\quad
\kappa_1 = \frac{2\pi k_\B T}{\hbar c}~. \label{casimirforce1d}
\end{eqnarray}
The primed sum symbol implies that the contribution of the zeroth
Matsubara pole at $n=0$ is counted for only one half (symbol written
here for a function $\varphi(n)$)
\begin{eqnarray}
\sum_n^\prime \varphi(n) \equiv \frac12 \varphi(0) +
\sum_{n=1}^\infty \varphi(n)~. \label{primedsum}
\end{eqnarray}
This final expression is always finite for any properly defined
model of mirrors.

\subsubsection*{Limiting cases}
In the limit $T\to0$, the ensemble of Matsubara poles becomes a cut
along the imaginary axis (the function $\displaystyle{\coth
\left(\hbar z/2k_\B T\right)}$ thus goes to +1 for $\Re z>0$ and to
-1 for $\Re z<0$). The discrete sum \eqref{casimirforce1d} is then
written as an integral over the positive part of the imaginary axis
\begin{eqnarray}
F_0= - \frac{\hbar c}{\pi} \int_0^\infty \mathrm{d}\kappa ~
\frac{\kappa ~ r [\imag\xi]} {e^{2\kappa L} - r [\imag\xi]}
\quad,\quad \kappa\equiv\frac{\xi}c ~. \label{casimirforce1dT0}
\end{eqnarray}

For perfect mirrors, i.e. when $r$ may be taken as unit value at all
frequencies contributing to the integral \eqref{casimirforce1dT0}, a
universal result is obtained, which no longer depends on the
specific properties of the mirrors
\begin{eqnarray}
F_0[r\to1] = - \frac{\hbar c}{\pi} \int_0^\infty \mathrm{d} \kappa ~
\frac{\kappa} {e^{2\kappa L} - 1} = - \frac{\hbar c \pi}{24L^2}~.
\label{casimirforce1dT0perfect}
\end{eqnarray}
We have used the fact that the Riemann zeta function
\begin{eqnarray}
\zeta(s) = \frac1{\Gamma(s)} \int_0^\infty \frac{x^{s-1}}{e^x-1} \,
\mathrm{d}x = \sum_{n=1}^\infty \frac1{n^s} ~.
\label{Riemannzetafunction}
\end{eqnarray}
takes the value $\zeta(2) = \pi^2/6 $ for $s=2$. Other values of the
same function will appear in various places in Casimir force
calculations with perfectly reflecting mirrors.

For real mirrors the product of reflection amplitudes $r=r_1r_2$
always has a modulus smaller than unity. It follows that the
integral is always regular with a smaller modulus than for perfect
mirrors. If the two mirrors are identical ($r_1=r_2$), then their
product $r$ is positive and the force is attractive as in the case
of perfect mirrors. The expression of the Casimir force as the
integral \eqref{casimirforce1dT0perfect} over imaginary frequencies
is convenient to discuss the meaning of the limit of perfect
mirrors. Taking as an example the model \eqref{ModelMirror}, we
indeed see that \eqref{casimirforce1dT0} tends to
\eqref{casimirforce1dT0perfect} as soon as the characteristic
frequency $\Omega$ at which the reflection falls down is larger than
the typical frequencies $\omega \sim c/L$ contributing to the
integral.

\subsection{The Casimir free energy and phase-shift interpretation}

We now write an expression for the Casimir free energy and show that
it can be given an interpretation in terms of scattering
phase-shifts. We also obtain expressions for the Casimir entropy and
Casimir internal energy.

\subsubsection*{The Casimir free energy}
We come back to the expression \eqref{casimirforce1df} of the force
as an integral over real frequencies and write it as the
differential of a free energy $\mathcal{F}$ with respect to $L$
\begin{eqnarray}
\label{casimirfreeenergy1d}
&&F=-\frac{\partial\mathcal{F}(L,T)}{\partial L} ~,\nonumber\\
&&\mathcal{F}= \int_0^\infty \frac{\mathrm{d}\omega}{2\pi}
\left(1+2\overline{n}_\omega\right) \frac\hbar{2\imath}
\ln\frac{d}{d^\ast} ~,\\
&&d\equiv 1-r e^{2ikL} ~. \nonumber
\end{eqnarray}
We show in the following that this formula can be given a nice
interpretation in terms of phase-shifts~\cite{Jaekel1991}. Note
that, though $\mathcal{F}$ could be changed by a $L-$independent
contribution without changing the result for $F$, the form of
$\mathcal{F}$ given in \eqref{casimirfreeenergy1d} is fixed by the
phase-shift interpretation discussed below.

\subsubsection*{The phase-shift interpretation}
\begin{figure}[tbh]
\includegraphics[width=0.45\textwidth]{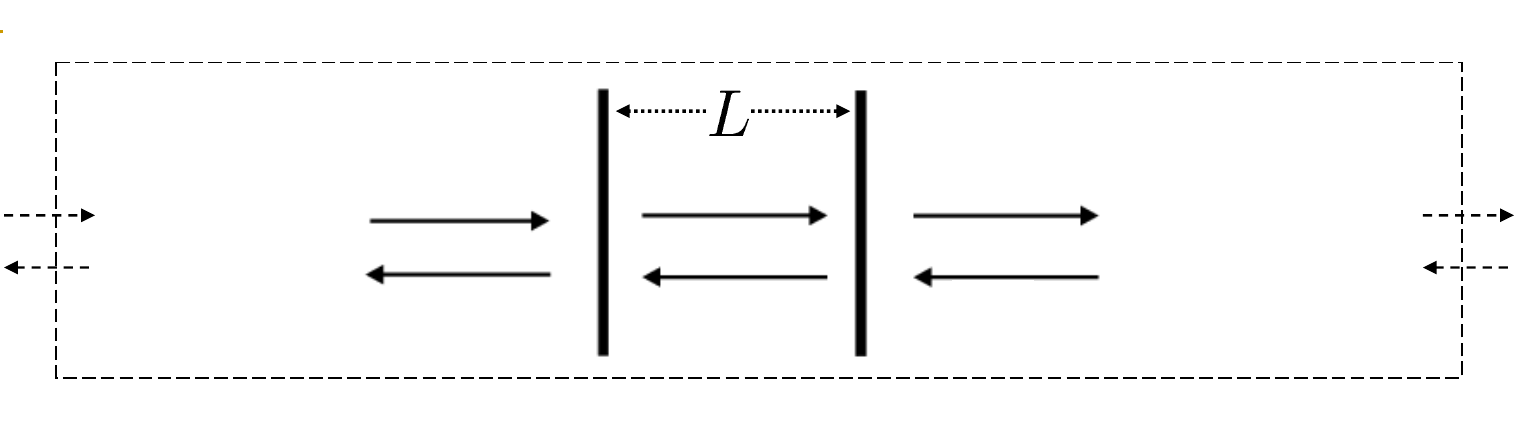}
\caption{The scattering system, a cavity of length $L$, is put in a
much larger quantization box, and the Casimir effect is calculated
as the change of free energy of the large box.}
\label{FigPhaseshift1d}
\end{figure}

To the aim of introducing the phase-shift interpretation, we put the
cavity of Figure \ref{FigCavity1d} inside a quantization box with a
much larger size $\calL \gg L$ and periodic conditions, as sketched
on Figure \ref{FigPhaseshift1d}.

In the absence of the scatterer, the modes in the large box would
have their wavelengths determined by the box size through $k_n^\pm
\calL = 2n\pi$, where $\pm$ labels the rightward and leftward
propagation directions which correspond to degenerate solutions. The
eigen-modes in the presence of the scatterer are thus determined by
the eigenvalues $e^{\imag\delta_n^\pm}$ of the unitary $S-$matrix
through $k_n^\pm \calL +\delta_n^\pm = 2n\pi$. The change of global
energy of the modes at a given frequency $\omega$ is then determined
by the quantity $-\hbar c \left(\delta_n^++\delta_n^-\right)$
proportional to the sum of the two phase-shifts at this frequency.
As a consequence, we do not need to solve the full eigenvalue
problem for the $S-$matrix, but we need only to calculate the
logarithm of its determinant $\ln\det S = \imag \left(\delta_n^+ +
\delta_n^-\right)$.

The relation between the expression \eqref{casimirfreeenergy1d} of
the free energy and the phase-shift interpretation is then fixed by
noting that the expression \eqref{globalSmatrix} written above for
the $S-$matrix associated to the cavity leads to the following
relation between the determinants
\begin{eqnarray}
\det S_{12}  = \left(\det S_1 \right) \left(\det S_2 \right) ~
\frac{d^\ast}{d} ~.
\end{eqnarray}
where we have denoted $S_{12}$ the scattering matrix for the
compound system consisting of the two mirrors 1 and 2 (it was simply
denoted $S$ in \eqref{globalSmatrix}). One thus deduces
\begin{eqnarray}
\ln\det S_{12}  = \ln\det S_1 + \ln\det S_2 + \ln \frac{d^\ast}{d}
~.
\end{eqnarray}

It is now clear that the Casimir free energy
\eqref{casimirfreeenergy1d} is given by a difference between changes
of free energies $\Delta\mathcal{F}$ calculated in three different
scattering configurations
\begin{eqnarray}
\mathcal{F}= \Delta\mathcal{F}_{12} - \Delta\mathcal{F}_{1} -
\Delta\mathcal{F}_{2} ~, \label{casimirfreeenergyDelta}
\end{eqnarray}
with each of these quantities determined by the phase-shifts for the
associated $S-$matrix
\begin{eqnarray}
\Delta\mathcal{F} _{12} = \int_0^\infty \frac{\dd\omega}{2\pi}
\left( 1+2\overline{n}_{\omega} \right) \frac{\imath\hbar}2
\ln\left(\mathrm{det} S_{12} \right) ~.
\label{casimirfreeenergyphaseshifts}
\end{eqnarray}
Similar expressions hold for the change of free energies
$\Delta\mathcal{F}_{1}$ and $\Delta\mathcal{F}_{2}$ associated
respectively with mirrors 1 and 2 placed in the large box on Figure
\ref{FigPhaseshift1d}.

In fact, each of the changes is itself a difference of free energies
calculated in the presence and in the absence of the scatterer
\begin{eqnarray}
\Delta\mathcal{F}_{12} \equiv \mathcal{F}_{12} - \mathcal{F}_{0}
\;,\; \Delta\mathcal{F}_{i} \equiv \mathcal{F}_{i} - \mathcal{F}_{0}
\;,\; i=1,2~,
 \label{casimirfreeenergyDiff3}
\end{eqnarray}
with the subscript 0 labeling the configuration with no scatterer.
In the end, the Casimir free energy is a difference involving four
different configurations
\begin{eqnarray}
\mathcal{F} &=& \left(\mathcal{F}_{12}-\mathcal{F}_{0} \right) -
\left(\mathcal{F}_{1} -\mathcal{F}_{0}\right)
 - \left(\mathcal{F}_{2} -\mathcal{F}_{0}\right) \nonumber\\
&=& \mathcal{F}_{12} - \mathcal{F}_{1}
 - \mathcal{F}_{2} + \mathcal{F}_{0} ~.
 \label{casimirfreeenergyDiff4}
\end{eqnarray}

It is worth emphasizing at this point that the definitions of
$\mathcal{F}_{12}$, $\mathcal{F}_{1}$, $\mathcal{F}_{2}$ and
$\mathcal{F}_{0}$ are affected by the problem of vacuum energy
discussed above. Meanwhile $\Delta\mathcal{F}_{12}$,
$\Delta\mathcal{F}_{1}$ and $\Delta\mathcal{F}_{2}$ are
self-energies, and their proper evaluation should involve some
renormalization procedure. In contrast, the Casimir free energy is
always a finite quantity, the evaluation of which does not require
renormalization~\cite{Jaekel1991}.

\subsubsection*{Casimir thermodynamics}
We can then write other thermodynamical functions associated with
the Casimir effect. One may in particular define an entropy
$\mathcal{S}$ from the free energy $\mathcal{F}$
\begin{eqnarray}
\mathcal{S}=-\frac{\partial\mathcal{F}(L,T)}{\partial T} ~,
 \label{entropy1d}
\end{eqnarray}
and then an internal energy $\mathcal{E}$
\begin{eqnarray}
\mathcal{E} = \mathcal{F} + T\mathcal{S} = \mathcal{F} - T
\frac{\partial{\mathcal{F}(L,T)}}{\partial T}~.
 \label{internalenergy1d}
\end{eqnarray}
These thermodynamical functions obey the usual thermodynamical
relations such as
\begin{eqnarray}
\mathrm{d}\mathcal{F}=-F\mathrm{d}L-\mathcal{S}\mathrm{d}T \quad,
\quad \mathrm{d}\mathcal{E}=-F\mathrm{d}L + T\mathrm{d}\mathcal{S}~.
\end{eqnarray}
The last relation just means that the change $\mathrm{d}\mathcal{E}$
of internal energy under a transformation is the sum of a mechanical
work $-F\mathrm{d}L$ and of a heat term $T\mathrm{d}\mathcal{S}$.

These thermodynamical functions can be deduced from the free energy
$\mathcal{F}(L,T)$ written previously as an integral over real
frequencies or, equivalently, from its form as a Matsubara sum
\begin{eqnarray}
\mathcal{F}(L,T) = k_\B T \sum_n^\prime \ln d[\imag \xi_n] ~.
\label{freeenergy1dmatsubara}
\end{eqnarray}
This expression can be obtained from \eqref{casimirfreeenergy1d} by
using anew the Cauchy's theorem. Equivalently, it can be obtained
from the Casimir force \eqref{casimirforce1d} written above as a sum
over Matsubara poles.

\section{The Casimir force in 3-dimensional space}
\label{Lecture3}

We come now to the last section of these lecture notes, which is
devoted to the discussion of the Casimir force in three-dimensional
(3d) space.

We will mainly discuss the geometry of two plane and parallel
mirrors, described by specular scattering amplitudes. This case
constitutes a trivial extension of the derivation in 1d space though
a few points have to be treated with greater care. As for the 1d
case, we will write a formula valid and regular at thermal
equilibrium at any temperature and for any optical model of mirrors
obeying causality and high frequency transparency
properties~\cite{Jaekel1991}. It reproduces the Casimir ideal
formula in the limits of perfect reflection and null temperature,
but can also be used for calculating the Casimir force between
arbitrary mirrors, as soon as the reflection amplitudes are
specified. For mirrors characterized by Fresnel reflection
amplitudes deduced from a linear and local dielectric function, the
Scattering Formalism leads to the same results as the Lifshitz's
method~\cite{Lifshitz1956,Dzyaloshinskii1961,Lifshitz1980}.

The section ends with a short presentation of the general
scattering formalism which allows one to deal with non specular
reflection and arbitrary geometries~\cite{Lambrecht2011}.

\subsection{Free electromagnetic fields in 3d space}

\subsubsection*{Modes for electromagnetic fields}
The free modes for electromagnetic fields in 3d space are the free
solutions of Maxwell equations in vacuum. They correspond to
frequency and wave-vector related through the dispersion relation
\begin{equation}
\frac{\omega ^{2}}{c^{2}}=k_{x}^{2}+k_{y}^{2}+k_{z}^{2} = \bk^{2} +
k_{z}^{2} \quad ,\quad \bk^{2} \equiv k_{x}^{2}+k_{y}^{2} ~.
\end{equation}
As mirrors will be specified later on to have their surfaces
parallel to the $\left(x,y\right)$ plane, $k_z$ is the longitudinal
part of the wave-vector and $\bk \equiv \left(k_x,k_y\right)$ its
transverse part. We introduce a direction of propagation as in 1d
calculations~: $\eta =+1$ corresponds to rightward propagation and
$\eta =-1$ to leftward propagation. $\eta$ is the sign of $k_z$
($\eta = \mathrm{sign} \left(k_z\right) = \mathrm{sign} \left(
\cos\theta \right)$) and it appears in the expression of $k_{z}$ in
terms of frequency and transverse wave-vector
$\displaystyle{k_z=\eta \sqrt{\frac{\omega^2}{c^2} - \bk^2}}$.

The propagation direction is defined by the incidence angle $\theta
$ and azimuth $\varphi $
\begin{eqnarray}
k=\frac\omega c \widehat k \quad,\quad \widehat k= \left(
\begin{array}{c}
\widehat k_x \\
\widehat k_y \\
\widehat k_z
\end{array}
\right)  = \left(
\begin{array}{c}
\sin\theta
\cos\varphi \\
\sin\theta
\sin\varphi \\
\cos\theta
\end{array}
\right) ~.
\end{eqnarray}
Unit wave-vector and polarization vectors for electric
($\widehat\alpha$) and magnetic ($\widehat\beta$) fields form an
ortho-normal basis for each polarization $p=$TE,TM with
$\widehat\beta^p = \widehat k \times \widehat\alpha^p$ in each case
\begin{eqnarray}
&&\widehat\alpha^\mathrm{TE} = \widehat\beta^\mathrm{TM} = \left(
\begin{array}{c}
-\sin\varphi \\
\cos\varphi \\
0
\end{array}
\right) ~, \nonumber\\ &&\widehat\alpha^\mathrm{TM} =
-\widehat\beta^\mathrm{TE} = \left(
\begin{array}{c}
\cos\theta
\cos\varphi \\
\cos\theta
\sin\varphi \\
-\sin\theta
\end{array}
\right) ~.
\end{eqnarray}

\subsubsection*{Electric and magnetic fields}
Electric and magnetic fields are then obtained as linear
superpositions of modes labeled by $m$ and $\eta$
\begin{eqnarray}
{E} = \underset {m,\eta} \sum
\sqrt{\frac{\hbar\omega}{2\varepsilon_0}} \widehat\alpha_{m,\eta}
&&\left(-\imag a_{m,\eta} e^{-\imag\omega t + \imag\bk\br +\imag k_z
z} \right. \nonumber\\  && \left.+\imag a_{m,\eta}^\dag
e^{\imag\omega t-\imag\bk\br -\imag k_z z}
\right) ~, \nonumber \\
{B} = \underset {m,\eta} \sum \sqrt{\frac{\hbar\omega\mu_0}{2}}
\widehat\beta_{m,\eta} &&\left(-\imag a_{m,\eta} e^{-\imag\omega
t+\imag\bk\br +\imag k_z z} \right. \nonumber\\ && \left. +\imag
a_{m,\eta}^\dag e^{\imag\omega t-\imag\bk\br -\imag k_z z} \right)
~.
\end{eqnarray}
$m$ gathers all quantum numbers but $\eta$ and can be written under
two alternative forms
\begin{eqnarray}
m \equiv \left(\bk,\vert k_z\vert,p\right) \quad \mathrm{or} \quad m
\equiv \left(\bk,\omega,p\right) ~.
\end{eqnarray}
$\eta$ is treated separately because the two modes $\eta=\pm1$
corresponding to the same $m$ will be coupled by the scattering
processes studied later on. The sum over modes has the following
definition ($A$ is an area introduced for defining integrals over
$\bk$)
\begin{eqnarray}
\underset m \sum &\equiv& \underset p \sum
\iint\frac{A\mathrm{d}^2\bk}{4\pi^2} \int\frac{\mathrm{d}\vert
k_z\vert}{2\pi} \nonumber\\ &=& \underset p \sum
\iint\frac{A\mathrm{d}^2\bk}{4\pi^2} \int\frac{\omega
\mathrm{d}\omega}{2\pi c^2 \vert k_z\vert}~. \label{sumovermodes}
\end{eqnarray}
Finally, the symbols $a_{m,\eta}$ and $a_{m,\eta} ^\dag$ appearing
in the positive and negative frequency components are the
annihilation and creation operators of quantum
electro-dynamics~\cite{Cohen-Tannoudji1989}. Their commutation
relations and correlations are immediate generalizations of the ones
\eqref{commutationrelations1D} and \eqref{correlationThermal}
written for 1d calculations.

Note that we have used the electromagnetic constants in vacuum
$\varepsilon _{0}$ and $\mu _{0}$. In the following, the symbol
$\varepsilon $ will be used as a relative permittivity with its
value in vacuum being unity and the relative permittivity will keep
its unit value in vacuum.

\subsubsection*{Energy density and radiation pressure}
Energy-momentum densities and radiation pressures are components of
the Maxwell stress tensor~\cite{Itzykson1985}. In particular, the
energy per unit volume $e$ is the component $T_{00}$
\begin{eqnarray}
e = \frac {\varepsilon_0 ~ {E}^\T{E}}2 + \frac {{B}^\T{B}}{2\mu_0}~,
\end{eqnarray}
where $\T$ symbolizes a matrix transposition. Substituting the
expression of free fields and calculating the mean value of energy
density in a thermal state, we obtain a simple generalization of the
expression \eqref{meanenergiesThermal} written for the 1d case
\begin{eqnarray}
\left\langle \, e\left(u\right) \,\right\rangle_\mathrm{therm} &=&
\underset{m,\eta} \sum {\frac{\hbar\omega_m}2}
\left(1+2\ovn_m\right) ~.
\end{eqnarray}

We come then to the radiation pressure on plane mirrors parallel to
the $\left(x,y\right)$ plane which corresponds to the component
$T_{zz}$ of the Maxwell stress tensor
\begin{eqnarray}
&&\frac{\varepsilon_0 \left( {E}_x{E}_x + {E}_y{E}_y - {E}_z{E}_z
\right) }2 \nonumber\\
&& + \frac{ {B}_x{B}_x + {B}_y{B}_y - {B}_z{B}_z } {2\mu_0} ~.
\end{eqnarray}
The quantum average of this quantity in a thermal state leads to
\begin{eqnarray}
\underset{m,\eta} \sum {\frac{\hbar\omega_m}2} \cos^2\theta_m
\left(1+2\ovn_m\right) ~. \label{radiationpressurein3d}
\end{eqnarray}
The sums over modes written in the preceding expressions are
infinite but the force on mirrors will be obtained as a finite
difference between the radiation pressure acting on their two sides.

\subsection{Scattering on plane mirrors in 3d space}

We consider plane mirrors parallel to the $\left(x,y\right)$ plane,
that is also surfaces of constant $z$ in 3d space. Scattering on
such mirrors is a simple extension of that described in the 1d case
because of symmetry considerations. Invariance under time
translations and lateral space translations implies that the
frequency frequency $\omega$, transverse wave-vector $\bk$ and
polarization $p$ are preserved. The scattering process is
\emph{specular} and it only affects the parameter $\eta$. We first
consider lossless mirrors and then address the problem of losses.

\subsubsection*{A lossless mirror in 3d space}
\begin{figure}[tbh]
\includegraphics[width=0.5\textwidth]{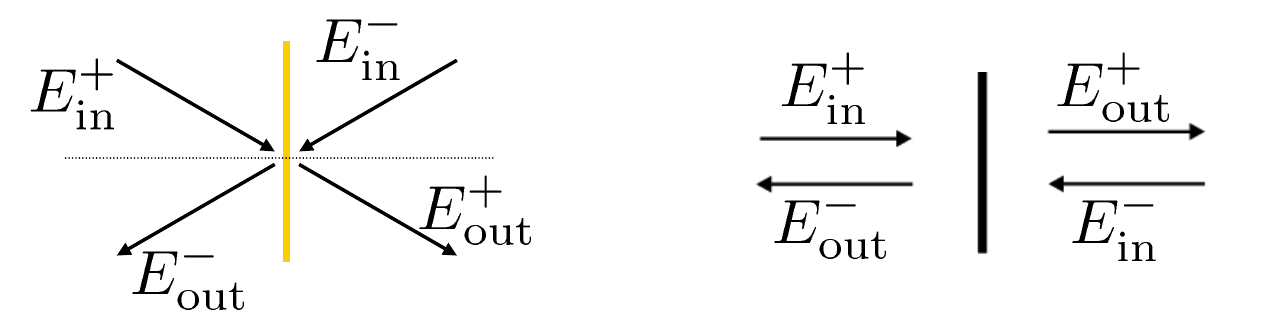}
\caption{Schematic representation of specular scattering upon a
lossless mirror in 3d space, with angles of incidence explicitly
shown on the left part but only implicit on the right part.}
\label{FigMirror3d}
\end{figure}

This specular scattering on a lossless mirror is schematically
represented on Figure \ref{FigMirror3d}. On the left part, the
angles of incidence are shown for the input and output fields. On
the right part in contrast, they are only implicit. It thus follows
that the sketch of the scattering process is alike that drawn on
Figure \ref{FigMirror1d} for the 1d case.

As in the 1d case, the scattering matrix is just a $2\times 2$
matrix giving the two output fields in terms of the two input ones
\begin{eqnarray}
&&\left(
\begin{array}{c}
{E}^+ _\out  \\
{E}^-_\out
\end{array}
\right)  = S_1 \left(
\begin{array}{c}
{E}^+ _\iin  \\
{E}^- _\iin
\end{array}
\right) ~, \nonumber\\ &&S_1 [m] = \left(
\begin{array}{cc}
t_1 [m] & r_1[m] e^{-2\imag\vert k_z\vert q_1} \\
r_1[m] e^{2\imag\vert k_z\vert q_1} & t_1[m]
\end{array}
\right)~.
\end{eqnarray}
Scattering amplitudes have been written as in \eqref{RealMirror}~:
$r_1$ and $t_1$ are defined for a mirror located at $x=0$ and depend
on the quantum number $m$ for real mirrors~; the general case is
then described by phases determined by the position $q_1$ of the
mirror and the longitudinal wave-vector $k_z$.

\subsubsection*{A lossless cavity in 3d space}
\begin{figure}[tbh]
\includegraphics[width=0.4\textwidth]{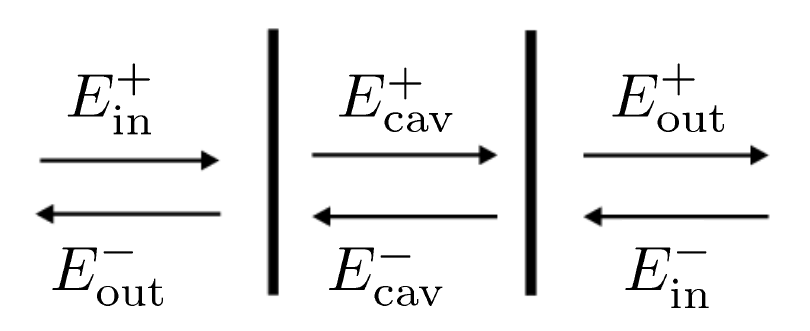}
\caption{Schematic representation of specular scattering upon a
cavity in 3d space, with the same convention as on the right part of
Figure \ref{FigMirror3d}.} \label{FigCavity3d}
\end{figure}

The specular scattering on a cavity made of two mirrors is
schematically represented on Figure \ref{FigCavity3d}.

Most calculations are the same as for the 1d case, with the effect
of the cavity described by a global scattering matrix and a
resonance matrix as in \eqref{SandRmatrices}
\begin{eqnarray}
&&\left(
\begin{array}{c}
E^+ _{\out} \\
E^-_{\out}
\end{array}
\right)  = S \left(
\begin{array}{c}
E^+ _{\iin}  \\
E^- _{\iin}
\end{array}
\right) ~, \nonumber\\ &&\left(
\begin{array}{c}
E^+ _{\cav} \\
E^-_{\cav}
\end{array}
\right)  = R \left(
\begin{array}{c}
E^+ _{\iin}  \\
E^- _{\iin}
\end{array}
\right) ~.
\end{eqnarray}
The matrices $S$ and $R$ are obtained from the scattering amplitudes
associated with the mirrors 1 and 2 as in the 1d case, that is to
say from equations \eqref{globalSmatrix} and \eqref{Rmatrix} with
the phase factors $e^{\pm\imag kL}$ replaced by $e^{\pm\imag \vert
k_z\vert L}$.

As the global $S-$matrix is unitary, radiation pressures on the
outer sides of the mirrors are given by the expression
\eqref{radiationpressurein3d}. Meanwhile, radiation pressures on the
inner sides of the mirrors are modified by a factor given by the
common value of the diagonal elements in $RR^\dag$
\begin{eqnarray}
&&g_m \equiv 1 + f_m + f_m^\ast = \frac{1-\vert{r_m}\vert^2} {\vert
1 - r_m  e^{2\imag\vert k_z\vert L} \vert^2} ~, \nonumber \\ &&f_m
\equiv \frac {r_m  e^{2\imag\vert k_z\vert L}}{1-r_m  e^{2\imag\vert
k_z\vert L}} ~. \label{gandf3d}
\end{eqnarray}
The effective pressure on the mirror is then given by the difference
between the pressures on its two sides
\begin{eqnarray}
&&\frac1A \underset m\sum \hbar\omega_m \cos^2\theta_m
\left(1+2\ovn_m\right)  \left( g_m - 1 \right) \nonumber\\
&&= \frac1A \underset m\sum \hbar\omega_m \cos^2\theta_m
\left(1+2\ovn_m\right) \left( f_m + f_m^\ast \right) ~.
\end{eqnarray}
The left-hand side corresponds to the radiation pressure
interpretation already discussed in the 1d case while the right-hand
side, written in terms of the causal function $f$, is the basis for
transformations using the Cauchy's theorem. However, some elements
are now to be treated with more care before the Casimir pressure is
obtained.

\subsection{The Casimir pressure between two plates in 3d space}

We now extend the derivations to take into account two  effects
which play an important role for calculations in 3d space. First, we
want to discuss the effect of dissipation inside the mirror and the
associated fluctuations. Then, we address the contributions of
evanescent waves which have to be added to that of ordinary
propagation waves~\cite{Genet2003}.

\subsubsection*{Generalization for lossy mirrors}
As already explained, the mirrors used in the experiments are made
of metals and they have losses associated to the dissipation inside
the matter. As a consequence, the scattering matrices defined above
cannot be considered as unitarity. In other words, there are
additional fluctuations accompanying losses which should be taken
into account in a large scattering matrix described the coupling of
the modes of interest and the noise
modes~\cite{Barnett1996,Barnett1998,Courty2000}.

At this point, it is worth using a theorem which gives the
commutators of the intra-cavity fields as the product of those well
known for fields outside the cavity and the factors $g_m$ introduced
in \eqref{gandf3d}. This theorem, which was demonstrated with an
increasing range of validity
in~\cite{Jaekel1991,Barnett1998,Genet2003}, proves that the
expression written in \eqref{gandf3d} for lossless mirrors is still
true for lossy mirrors. We then assume thermal equilibrium for the
whole system, which means that input fields as well as fluctuations
associated with electrons, phonons and any loss mechanism inside the
mirrors have to correspond to the same temperature $T$, whatever
their microscopic origin may be. Then, the radiation pressures are
given by the expressions written above and the last equation becomes
the Casimir pressure, valid for lossy as well as lossless mirrors,
\begin{eqnarray}
P = \frac1A \underset m\sum \hbar\omega_m \cos^2\theta_m
\left(1+2\ovn_m\right) \left( f_m + f_m^\ast \right) ~.
\label{pressure3dff}
\end{eqnarray}

\subsubsection*{Generalization for evanescent waves}
Up to now, we have discussed the contributions of ordinary waves
freely propagating outside and inside the cavity, which correspond
to real wave-vectors with $k_z$ real, that is also $\omega > c
\vert\bk\vert$. Equation \eqref{pressure3dff} thus reflects the
intuitive picture of radiation pressure of vacuum and thermal
fluctuations, as discussed above in the 1d case. In the 3d case, we
have also to take into account the contribution of evanescent waves
(see \S 1.5.4 in~\cite{Born1999}),
which correspond to imaginary values of $k_z$, that is also
frequencies in the interval $0< \omega < c \vert\bk\vert$.

Those waves propagate inside the mirrors with an incidence angle
larger than the limit angle and they also exert a radiation pressure
on the mirrors, due to the frustrated reflection phenomenon. In
fact, the expression \eqref{pressure3dff} of the Casimir pressure
has to be understood as including these
contributions~\cite{Genet2003}. They are obtained through an
analytical continuation of those of ordinary waves, using the well
defined analytic behavior of the causal function $f_m$. This
analytical continuation has to be defined for fixed values of the
lateral wave-vector $\bk$ and polarization $p$.

Using the definition \eqref{sumovermodes} of the sum over modes and
the relation $\cos\theta=ck_z/\omega$, one finally gets the Casimir
pressure \eqref{pressure3dff} as the real part of an integral over
the positive part $\mathbb{R}^+$ of the real axis (to be compared
with \eqref{casimirforce1df})
\begin{eqnarray}
\label{pressure3df} &&P = P_r + P_r^\ast ~, \\ &&P_r \equiv \sum_p
\iint\frac{\mathrm{d}^2\bk}{4\pi^2} {\int}_0^\infty
\frac{\mathrm{d}\omega}{2\pi} \hbar \vert k_z\vert f_m ~
\coth\frac{\hbar\omega}{2k_\B T} ~. \nonumber
\end{eqnarray}
We stress again that the integral over $\mathbb{R}^+$ includes the
contributions of evanescent waves $0<\omega<c\vert\bk\vert$ besides
those of ordinary waves $c\vert\bk\vert<\omega$.

\subsubsection*{Casimir pressure as an integral over imaginary frequencies}
One now transforms the expression \eqref{pressure3df} of the Casimir
pressure by using Cauchy's theorem as in the 1d case. In the end of
the derivation, the Casimir pressure is the integral over a contour
which encircles the imaginary axis, and it is thus found to be a
discrete sum of the values of the function $\vert k_z\vert f_m$ at
the Matsubara poles
\begin{eqnarray}
P=-k_\mathrm{B}T \sum_p \int \frac{\mathrm{d}^2\mathbf{k}}{(2\pi)^2}
\sum_n^\prime \frac{2\kappa_n\, r_\mathbf{k}^p[i\xi_n] e^{-2\kappa_n
L}} {1-r_\mathbf{k}^p[i\xi_n] e^{-2\kappa_n L}} ~,
\label{pressure3d}
\end{eqnarray}
where $\kappa$ is obtained from $\vert k_z\vert$ through an
analytical continuation for imaginary frequencies $\omega_n= i\xi_n$
\begin{eqnarray}
\kappa_n=\sqrt{\bk^2+\frac{\xi_n^2}{c^2}} \quad,\quad \xi_n=n
\frac{2\pi k_\mathrm{B}T}{\hbar} ~.
\end{eqnarray}
As in the 1d case, this expression is finite for any properly
defined model of mirrors.

\subsubsection*{Zero temperature limit}
In the limit $T\to0$, the ensemble of Matsubara poles becomes a cut
along the imaginary axis and the Matsubara sum \eqref{pressure3d}
becomes an integral over the positive part of the imaginary axis
($P_0\equiv P_{T=0}$)
\begin{eqnarray}
P_0 &=& -2 \underset p\sum
\iint\frac{\mathrm{d}^2\mathbf{k}}{4\pi^2}
\int_0^\infty\frac{\mathrm{d}\xi}{2\pi} \hbar\kappa
\frac{r_\mathbf{k}^p[i\xi] e^{-2\kappa L}} {1-r_\mathbf{k}^p[i\xi]
e^{-2\kappa L}} ~,\nonumber\\
&&\kappa=\sqrt{\bk^2+\frac{\xi^2}{c^2}} ~. \label{pressure3dT0}
\end{eqnarray}

For perfect mirrors, i.e. when $r$ may be taken as unit value at all
frequencies contributing to the integral \eqref{pressure3dT0}, a
universal result is obtained, which no longer depends on the
specific properties of the mirrors
\begin{eqnarray}
P_0[r\to1] = -\frac{\hbar c}{\pi^2} \int_0^\infty \frac{\kappa^3
\mathrm{d}\kappa} {e^{2\kappa L}-1} = -\frac{\hbar c\pi^2}{240L^4}
~. \label{pressure3dT0perfect}
\end{eqnarray}
We have used the fact that the Riemann zeta function
\eqref{Riemannzetafunction} takes the value $\zeta(4)=\pi^4/90$ for
$s=4$.

\subsection{Models for metallic mirrors}

We now come back to the discussion presented in section \ref{Lecture1}
for the models of metallic mirrors used in the experiments. The common
model is that of bulk mirrors, in fact thick slabs, made with metals
for example gold. The optical response of the metal is described
by a local dielectric response function and the reflection amplitudes
on each mirror then deduced by using Fresnel laws.

\subsubsection*{Fresnel reflection amplitudes}
Reflection amplitudes on a thick slab described by a local
dielectric response $\varepsilon$ are given by the well known
Fresnel laws for the TE and TM polarizations (see \S 1.5.1
in~\cite{Born1999} or \S 86 in~\cite{Landau1980})
\begin{eqnarray}
r_1^\mathrm{TE}[\omega]=\frac{k_z-K_z}{k_z+K_z} \quad,\quad
r_1^\mathrm{TM}[\omega] =\frac{K_z-\varepsilon k_z} {K_z+\varepsilon
k_z} ~,
\end{eqnarray}
where $K_z$ and $k_z$ are the longitudinal wave-vectors
in matter and vacuum respectively
\begin{eqnarray}
K_z = \sqrt{\varepsilon\frac{\omega ^2}{c^2}-\bk^2} \quad,\quad k_z =
\sqrt{\frac{\omega ^2}{c^2}-\bk^2} ~.
\end{eqnarray}
When the frequency is continued to imaginary values,
these expressions become
\begin{eqnarray}
\label{fresnellaws}
&&r_1^\mathrm{TE}[i\xi]=\frac{\kappa-K}{\kappa+K}
\quad,\quad
r_1^\mathrm{TM}[i\xi] =\frac{K-\varepsilon[i\xi] \kappa}
{K+\varepsilon[i\xi] \kappa} ~, \nonumber\\
&&K = \sqrt{\varepsilon[i\xi] \frac{\xi ^2}{c^2}+k^2} \quad,\quad
\kappa = \sqrt{\frac{\xi ^2}{c^2}+k^2} ~.
\end{eqnarray}

When the reflection amplitudes \eqref{fresnellaws} are inserted
in the expression \eqref{pressure3d} of the pressure,
the formula obtained in 1956 by E.M. Lifshitz~\cite{Lifshitz1956}
and derived again in 1961 by I.E. Dzyaloshinskii, E.M. Lifshitz and L.P.
Pitaevskii~\cite{Dzyaloshinskii1961} is recovered (see
also Chap. VIII in~\cite{Lifshitz1980}).
We may stress at this point that this formula was obtained
through a different derivation, with all fluctuations
originating from matter. In fact this expression
was written in terms of the dielectric function and not
of reflection amplitudes. To our best knowledge,
Kats was the first to notice that it could be
written in terms of the reflection amplitudes~\cite{Kats1977}.
We may quote at this point a few alternative derivations
of this formula through different
methods~\cite{Langbein1970,Renne1971,Candelas1982,Kupiszewska1990,%
Scheel2008,Kenneth2008,Rahi2009,Intravaia2012,Reid2013}.

We may remark that the optical response of the bulk material cannot
always be described by a local dielectric function.
In this case, the description in terms of reflection amplitudes is
still valid, though the reflection amplitudes cannot be written
under the specific forms \eqref{fresnellaws}.
More detailed descriptions can for example be determined from
microscopic models of metallic conduction.
Attempts in this direction and discussions can be
found for example
in~\cite{Pitaevskii2008prl,Dalvit2008prl,Svetovoy2008prl,%
Geyer2009prl,Pitaevskii2009prl,Decca2009prl,Dalvit2009prl}.

\subsubsection*{Models for dielectric functions}
As already discussed in section \ref{Lecture1}, the dielectric
function $\varepsilon$ is usually obtained from optical
data~\cite{Lambrecht2000,Svetovoy2008prb}. These data are then
extrapolated at low frequencies by using the dissipative Drude model
\eqref{Drudemodel} for the conductivity of the
metal~\cite{Ashcroft1976}. In contrast to the plasma model
\eqref{plasmamodel} which corresponds to the lossless limit $\gamma
= 0$, the dissipative Drude model meets the well-known fact that
gold has a finite static conductivity \eqref{finiteconductivity}.

The two models lead to different predictions for the Casimir pressure,
in particular at the limits of large distances or large temperatures.
This can be attributed to different limiting cases at zero frequency.
The Drude model indeed corresponds to
\begin{eqnarray}
\underset {\omega\to0} \lim r_\bk^\mathrm{TE}[\omega] = 0 \;,\;
\underset {\omega\to0} \lim r_\bk^\mathrm{TM}[\omega] = 1 \;,\;
\gamma\neq0 ~,
\end{eqnarray}
whereas the plasma model leads to
\begin{eqnarray}
\underset {\omega\to0} \lim r_\bk^\mathrm{TE}[\omega] = 1 \;,\;
\underset {\omega\to0} \lim r_\bk^\mathrm{TM}[\omega] = 1 \;,\;
\gamma=0 ~.
\end{eqnarray}

\subsubsection*{High temperature limit}
At the limit of high temperatures $\kappa_1 L \gg 1$, the Matsubara
poles with $n>0$ give an exponentially small contribution to the
Casimir pressure \eqref{pressure3d}. The result is thus dominated by
the contribution of the zeroth pole $n=0$ ($P_\infty\equiv
P_{T\to\infty}$)
\begin{eqnarray}
P_\infty = -k_\B T \sum_p \int \frac{\mathrm{d}^2\bk}{(2\pi)^2}
\frac{\vert\bk\vert r_\bk^p[0] e^{-2 \vert\bk\vert L}} {1-r_\bk^p[0]
e^{-2 \vert\bk\vert L}} ~. \label{pressure3dinf}
\end{eqnarray}

It follows that the asymptotic Casimir pressure \eqref{pressure3dinf}
has different values for a dissipative model
\begin{eqnarray}
P_\infty = -\frac{k_\mathrm{B}T}{2\pi} \int_0^\infty \frac{k^2\,
\mathrm{d}{k}} {e^{2k L}- 1} = -\frac{k_\mathrm{B}T}{8\pi L^3}
\zeta(3) \;,\; \gamma\neq0
\end{eqnarray}
and a lossless model
\begin{eqnarray}
P_\infty = -\frac{k_\mathrm{B}T}{\pi} \int_0^\infty \frac{k^2\,
\mathrm{d}{k}} {e^{2k L}- 1} = -\frac{k_\mathrm{B}T}{4\pi L^3}
\zeta(3) \;,\; \gamma=0 .
\end{eqnarray}
$\zeta(3) \simeq 1.202$ is the value of the Riemann zeta function
\eqref{Riemannzetafunction} for $s=3$.
As already discussed in section \ref{Lecture1}, there is a large
factor 2 between the two predictions because the two polarizations
contribute equally for the plasma model, as well as for perfect
mirrors, whereas only one polarization contributes for the Drude model.

\subsection{The Casimir free energy in 3d space}

Using the results already obtained in the 1d case,
we write an expression for the Casimir free energy and show that
it can be given an interpretation in terms of scattering
phase-shifts. We also obtain expressions for the Casimir entropy and
Casimir internal energy.

\subsubsection*{The Casimir free energy and phase-shift interpretation}
The expression \eqref{pressure3df} of the Casimir pressure
can be written as the differential $P=-\partial\mathcal{F}/\partial V$
of a free energy $\mathcal{F}(V,T)$ with respect to the volume $V\equiv AL$
\begin{eqnarray}
\mathcal{F} &=& \sum_p A \iint\frac{\mathrm{d}^2\bk}{4\pi^2}
\int_0^\infty \frac{\mathrm{d}\omega}{2\pi}
\left(1+2\overline{n}_\omega\right) \nonumber\\ && \times
\frac\hbar{2\imath} \ln\frac{1-r e^{2i \vert k_z\vert L}}{1-r^\ast
e^{-2i \vert k_z\vert L}} ~. \label{casimirfreeenergy3d}
\end{eqnarray}
The interpretation of this formula in terms of phase-shifts is the
same as in the 1d case~\cite{Jaekel1991} and we do not repeat here
the derivation presented in section \ref{Lecture2}. In the end of
this derivation, the Casimir free energy \eqref{casimirfreeenergy3d}
is given by a difference \eqref{casimirfreeenergyDelta} of changes
of free energies calculated in three different scattering
configurations, with each of these quantities determined by an
integral \eqref{casimirfreeenergyphaseshifts} over all modes. As
already emphasized, the Casimir free energy is always a finite
quantity, the evaluation of which does not require a
renormalization~\cite{Jaekel1991}.

\subsubsection*{Casimir thermodynamics}
We can then write other thermodynamical functions associated with
the Casimir effect. One may in particular define an entropy
$\mathcal{S}=-\partial\mathcal{F}/\partial T$
and then an internal energy $\mathcal{E} = \mathcal{F} + T\mathcal{S}$.
These thermodynamical functions obey the usual thermodynamical
relations ($V\equiv AL$)
\begin{eqnarray}
\mathrm{d}\mathcal{F}=-P\mathrm{d}V-\mathcal{S}\mathrm{d}T \quad,
\quad \mathrm{d}\mathcal{E}=-P\mathrm{d}V + T\mathrm{d}\mathcal{S}~.
\end{eqnarray}

These thermodynamical functions can be deduced from the free energy
\eqref{casimirfreeenergy3d} written previously as an integral over real
frequencies or, equivalently, from its form as a Matsubara sum
\begin{eqnarray}
\mathcal{F}=k_\mathrm{B}T
\sum_n^\prime \mathrm{Tr} \ln d[i\xi_n] ~.
\label{casimirfreeenergy3dmatsubara}
\end{eqnarray}
We have used the denominator of the $S-$matrix to write the free energy
\begin{eqnarray}
d_\mathbf{k}^p[i\xi_n] = 1-r_\mathbf{k}^p[i\xi_n] e^{-2\kappa_n L}
\;,\; \kappa_n = \sqrt{\frac{\xi_n ^2}{c^2}+k^2} ~,
\label{denominator3dmatsubara}
\end{eqnarray}
and we have also introduced a trace over polarizations
and transverse wave-vectors in order to simplify the expression
\begin{eqnarray}
\Tr d \equiv
\sum_p ~\int \frac{A\mathrm{d}^2\mathbf{k}}{(2\pi)^2}
~ d_\mathbf{k}^p ~.
\label{tracedef}
\end{eqnarray}
The expression \eqref{casimirfreeenergy3dmatsubara} for the
free energy is completely equivalent to the expression
\eqref{pressure3d} written above for the pressure.

\subsection{General scattering formula}

We now present a more general scattering formula allowing one to
calculate the Casimir force between stationary disjoint objects with
arbitrary shapes. We also review rapidly some of the applications of
this new method which have been dedicated in the recent years to the
study of non trivial geometries. More exhaustive reviews of the
topic can be found
in~\cite{Lambrecht2006,Emig2007,Milton2008,Lambrecht2011,Rahi2011,Rodriguez2011}.

\subsubsection*{The non-specular scattering formula}
\begin{figure}[tbh]
\includegraphics[width=0.4\textwidth]{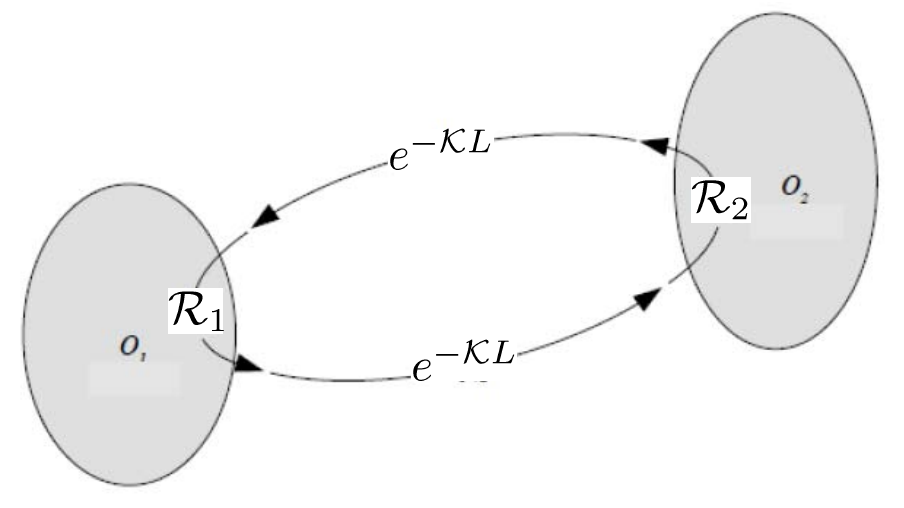}
\caption{General scattering configuration with
two objects $O_1$ and $O_2$ in vacuum~:
the Casimir free energy can be calculated from the
denominator~\eqref{denominator3dns} describing
the resonance properties of the cavity formed by the two
objects and given
by the reflection matrices $\calR_1$ and $\calR_2$
and the propagation matrices $e^{-\calK L}$.}
\label{FigScattering}
\end{figure}

We consider a geometrical configuration with two disjoint scatterers
at rest in electromagnetic vacuum (Fig. \ref{FigScattering}). 
The scattering matrix is now a
large matrix which accounts for non-specular reflection mixing
different wave-vectors and polarizations while preserving frequency
if the scatterers are at rest. Of course, the non-specular
scattering formula is the generic one while specular reflection is
an idealization for perfectly plane and flat plates.

The reasoning presented above leads to an expression of the Casimir
free energy as the sum of all phase-shifts contained in the large
$S-$matrix~\cite{Lambrecht2006}. Causality and high-frequency
transparency then allow one to write it as a Matsubara sum (to be
compared with \eqref{casimirfreeenergy3dmatsubara})
\begin{eqnarray}
\label{casimirfreeenergy3dns} \calF = k_\B T \sum_m^\prime \Tr \ln
\calD (i\xi_m) \;,\quad \xi_m \equiv \frac{2\pi m k_\B T}\hbar ~.
\end{eqnarray}
Each term in this sum is the trace of a large matrix which describes
all couplings at a given frequency. The matrix $\calD$ (here written
at Matsubara frequencies $\omega_m=i\xi_m$) is the denominator of
the scattering matrix. It describes the resonance properties of the
cavity formed by the two objects (to be compared with
\eqref{denominator3dmatsubara})
\begin{eqnarray}
\calD= 1 - \calR_1 e^{ -\calK L } \calR_2 e^{ -\calK L } ~.
\label{denominator3dns}
\end{eqnarray}

The matrices $\calR_1$ and $\calR_2$ represent reflection on the two
objects 1 and 2 respectively while $e^{-\calK L}$ describes
propagation. Note that the matrices $\calD$, $\calR_1$ and
$\calR_2$, which were diagonal in the plane wave basis for specular
scattering, are no longer diagonal in the general case of non
specular scattering. The propagation factors remain diagonal in this
basis with their eigenvalues $\kappa$ written as in
(\ref{denominator3dmatsubara}). Note that the expression
(\ref{casimirfreeenergy3dns}) does not depend on the choice of a
specific basis.

\subsubsection*{Applications to non trivial geometries}
The fact that geometry plays a non trivial role in the context of
vacuum energy and Casimir forces has been noticed for a long
time~\cite{Balian1977,Balian1978}. In the plane-sphere geometry for
example, the \emph{Proximity Force Approximation} (PFA) can only be
valid when the radius is much larger than the
separation~\cite{Schaden2000,Jaffe2004,Schroder2005}. Recently,
calculations have used the general non-specular scattering formula
in order to push the theory beyond the PFA and thus open roads to a
new
domain~\cite{Lambrecht2006,Emig2007,Milton2008,Lambrecht2011,Rahi2011,Rodriguez2011}.

The first application of the non-specular scattering formula to
calculations beyond the PFA was developed
in~\cite{Neto2005epl,Neto2005pra} to study the roughness correction
to the Casimir force between two planes, in a perturbative expansion
with respect to the roughness amplitude. Another geometry studied
with this method is that of surfaces with periodic corrugations. As
lateral translation symmetry is broken, the Casimir force contains a
lateral component which is smaller than the normal one, but has
nevertheless been measured in dedicated experiments~\cite{Chen2002}.
Calculations beyond the PFA have first been performed with the
simplifying assumptions of perfect reflection~\cite{Buscher2005} or
shallow corrugations~\cite{Rodrigues2006prl,Rodrigues2007pra}. As
could be expected, the PFA was found to be accurate only in the
limit of large corrugation wavelengths.

In recent years, experiments have been able to probe the beyond-PFA
regime~\cite{Chan2008,Chiu2009,Intravaia2013}. Exact calculations of
the forces between real mirrors with deep corrugations have been
performed~\cite{Lambrecht2008prl,Lambrecht2008nat} and comparisons
between theory and experiments presented~\cite{Bao2010,Guerout2013}.
The same kind of calculations was performed for studying the Casimir
torque which appears between two corrugated surfaces with non
aligned corrugations~\cite{Rodrigues2006epl}.

\subsubsection*{The plane-sphere geometry}
Another important application corresponds to the plane-sphere
geometry used in most Casimir force experiments and for which
explicit \emph{exact calculations} (see a discussion of the meaning
of this expression below) have recently been developed (see
discussions of the various methods devoted to this problem
in~\cite{Reynaud2008jpa,Emig2008jpa,Bordag2008jpa,Wirzba2008jpa,%
Klingmuller2008jpa}).

Indeed, exact multipolar expansions have been written for the force
in the plane-sphere geometry~\cite{Neto2008pra,Emig2008jsm}. These
calculations have now been performed for metallic surfaces coupled
to electromagnetic vacuum, at zero~\cite{Canaguier-Durand2009prl} or
non zero
temperature~\cite{Canaguier-Durand2010prl,Canaguier-Durand2010pra},
and this has opened the way to a comparison with theory of the only
experimental study devoted to a test of PFA in the plane-sphere
geometry~\cite{Krause2007prl}. The results of these calculations is
often presented in terms of a ratio $\rho$ of the exact result to
the PFA approximation, which is close to unity when the aspect ratio
$x\equiv L/R$ is small. An alternative representation is in terms of
the \emph{slope} $\beta$ defined by $\rho\equiv1+x\beta$.

In fact, the \emph{exact result} is an infinite series corresponding
to a multipolar expansion over the numbers $\left(\ell, m\right)$
which label the basis of spherical waves (with $\vert m \vert \leq
\ell$). Sums have to be truncated for the numerics to
$\ell\le\ell_\max$ and the precision is thus limited for small
values of the aspect ratio. The minimum aspect ratio for precise
calculations scales as the inverse of the maximum multipolar index
$\ell_\max$. For some time, extrapolations at low values of $x$ were
used to obtain bounds useful for the values of experimental
interest. Such a procedure used the idea that the slope $\beta$
could be represented as a Taylor expansion of the aspect
ratio~\cite{Fosco2011prd,Bimonte2011apl,Bimonte2012epl}. However
this idea was finally shown to be wrong in the high temperature
limit which corresponds to a classical Casimir effect with an
entropic origin~\cite{Canaguier-Durand2012pra,Bimonte2012prl}. As a
consequence, the question of accurate evaluations for the deviation
from PFA remains open.

\subsubsection*{Nano-spheres and atoms close to surfaces}
We will end up this paper by considering another case of interest,
that of dielectric nano-spheres above a plane, which tends
at the limit of a very small sphere to the Casimir-Polder problem.

The calculations easier for nano-spheres than for large spheres
because a small number of multipoles is sufficient for reaching a
good accuracy. An example of such calculations is given
in~\cite{Canaguier-Durand2011pra} for nano-diamonds above a copper
plate. Of course, the Casimir-Polder expression between an atom and
a plane is recovered in the limit of large distances $R\gg L$ where
the dipole approximation is sufficient. The Casimir-Polder limit is
written in terms of a polarizability for the atom (expression given
in~\cite{Messina2009}). The polarizability of a small sphere is that
of a large atom as it is proportional to the volume of the
sphere~\cite{Canaguier-Durand2011pra}.

\subsubsection*{Conclusion}
The general scattering formula allows one to describe
all long-range interactions in a unified manner as the change of
vacuum energy due to the presence of scatterers in vacuum.
Atoms as well as mirrors are completely characterized by
scattering amplitudes which have of course different properties.
In particular, atomic scattering is weak so that perturbation theory
is in general sufficient whereas scattering by mirrors may be
saturated (up to 100\% reflection).
Atoms are local probes of vacuum whereas mirrors are not.

The same general scattering formula thus allows one to treat different
cases which lead to a variety of phenomena~\cite{Lambrecht2006},
for example atom/atom, atom/plate or plate/plate interactions,
with plane or spherical plates as well as nano-structured surfaces.
In order to quote a few examples, it has been possible to
study the Casimir-Polder forces or torques for atoms near corrugated
surfaces~\cite{Dalvit2008,ContrerasReyes2010,Impens2010}.

\subsection*{Acknowledgments} The authors thank R.O. Behunin, A.
Canaguier-Durand, I. Cavero-Pelaez, H.B. Chan, D.A.R. Dalvit, R.S.
Decca, G. Dufour, T. Ebbesen, E. Fischbach, C. Genet, R. Gu\'erout,
G.-L. Ingold, F. Intravaia, M.-T. Jaekel, A. Liscio, J. Lussange,
P.A. Maia Neto, K.A. Milton, V.V. Nesvizhevsky, G. Palasantzas, P.
Samori, C. Speake, A. Voronin and Y. Zeng for contributions or
discussions related to this paper, and the ESF Research Networking
Programme CASIMIR (www.casimir-network.com) for providing excellent
opportunities for discussions on the Casimir effect and related
topics.


\newcommand{\REVIEW}[4]{\textit{#1} \textbf{#2} #3 (#4)}
\newcommand{\Review}[1]{\textit{#1}}
\newcommand{\Volume}[1]{\textbf{#1}}
\newcommand{\Book}[1]{\textit{#1}}
\newcommand{\Eprint}[1]{\textsf{#1}}
\def\etal{\textit{et al}}

\end{document}